\documentclass[12pt,preprint]{aastex}
\usepackage[version=3]{mhchem}
\usepackage{multirow}

\makeatletter
\newcounter{reaction}
\renewcommand\thereaction{C\,\arabic{reaction}}
\newcommand\reactiontag{\refstepcounter{reaction}\tag{\thereaction}}
\newcommand\reaction@[2][]{\begin{equation}\ce{#2}%
\ifx\@empty#1\@empty\else\label{#1}\fi%
\reactiontag\end{equation}}
\newcommand\reaction@nonumber[1]{\begin{equation*}\ce{#1}%
\end{equation*}}
\newcommand\reaction{\@ifstar{\reaction@nonumber}{\reaction@}}
\makeatother

\shorttitle{Surface Composition of Rocky Exoplanets}
\shortauthors{Hu et al.}

\begin{document}

\title{Theoretical Spectra of Terrestrial Exoplanet Surfaces}

\author{Renyu Hu}
\affil{Department of Earth, Atmospheric and Planetary Sciences, Massachusetts Institute of Technology, Cambridge, MA 02139}
\email{hury@mit.edu}

\author{Bethany L. Ehlmann}
\affil{Division of Geological and Planetary Sciences, California Institute of Technology, Pasadena, CA, 91125 \\
Jet Propulsion Laboratory, Pasadena, CA, 91109}

\and

\author{Sara Seager}
\affil{Department of Earth, Atmospheric and Planetary Sciences, Massachusetts Institute of Technology, Cambridge, MA 02139}

\begin{abstract}
We investigate spectra of airless rocky exoplanets with a theoretical framework that self-consistently treats reflection and thermal emission.
We find that a silicate surface on an exoplanet is spectroscopically detectable via prominent Si-O features in the thermal emission bands of 7 - 13 $\mu$m and 15 - 25 $\mu$m. The variation of brightness temperature due to the silicate features can be up to 20 K for an airless Earth analog, and the silicate features are wide enough to be distinguished from atmospheric features with relatively high-resolution spectra. The surface characterization thus provides a method to unambiguously identify a rocky exoplanet.
Furthermore, identification of specific rocky surface types is possible with the planet's reflectance spectrum in near-infrared broad bands.
A key parameter to observe is the difference between K band and J band geometric albedos ($A_{\rm g}({\rm K})-A_{\rm g}({\rm J})$): $A_{\rm g}({\rm K})-A_{\rm g}({\rm J})>0.2$ indicates that more than half of the planet's surface has abundant mafic minerals, such as olivine and pyroxene, in other words primary crust from a magma ocean or high-temperature lavas; $A_{\rm g}({\rm K})-A_{\rm g}({\rm J})<-0.09$ indicates that more than half of the planet's surface is covered or partially covered by water ice or hydrated silicates, implying extant or past water on its surface.
Also, surface water ice can be specifically distinguished by an H-band geometric albedo lower than the J-band geometric albedo. The surface features can be distinguished from possible atmospheric features with molecule identification of atmospheric species by transmission spectroscopy.
We therefore propose that mid-infrared spectroscopy of exoplanets may detect rocky surfaces, and near-infrared spectrophotometry may identify ultramafic surfaces, hydrated surfaces and water ice.
\end{abstract}


\section{Introduction}

Rocky exoplanets have been discovered by transit surveys. A primary transit happens when a planet passes in front of its host star and blocks a part of the star, so that the depth of the transit corresponds to the ratio between the planetary radius and the stellar radius. The planet's radius together with its mass provides the mean density and clues on its interior structure (See Rogers \& Seager 2010 and references therein). A few exoplanets have been suggested to have rocky surfaces, because the mass and radius constraints indicate that the planets are predominantly rocky with no extensive atmosphere envelope, i.e., likely rocky surfaces. Corot-7b opens up the possibility of close-in airless rocky exoplanets (L\'eger et al. 2009; Queloz et al. 2009). Due to the small star-planet distance, Corot-7b may have molten or even vaporized metals on its sub-stellar surface (L\'eger et al. 2011). Recently observations of transits of 55 Cnc e, a 8-$M_{\earth}$ planet around a G8V star, determine the planetary radius to be 2.1 $R_{\earth}$, which suggests that it can be a rocky planet (Winn et al., 2011; Demory et al., 2011). The Kepler mission, with unprecedented photometric precision,  is very powerful in discovering small-size transiting exoplanets. Kepler-10b, a 4.5-$M_{\earth}$ planet, is the first rocky exoplanet discovered by Kepler (Batalha et al. 2011). Kepler recently discovered several planets with size in the``super earth" regime, including Kepler-11b (Lissauer et al. 2011), Kepler-18b (Cochran et al. 2012), Kepler-20b (Gautier et al. 2012) and notably Kepler-22b in its host star's habitable zone (Borucki et al. 2012). Due to uncertainties of the planets' radii and masses, however, the composition of these Kepler super-Earths is ambiguous. They can be predominantly rocky like Earth, or have significant gas envelope like Neptune. Also, Kepler has detected Earth-sized transiting planets, Kepler-20e and Kepler-20f, with no constraints on their masses due to difficulties of followup radial-velocity observations (Fressin et al. 2012). With the progress of the Kepler mission, other transit surveys and followup observations, more and more exoplanets that potentially have rocky surfaces will be discovered and confirmed.

The purpose of this paper is to identify mineral-specific spectral features that would allow characterization of the surface composition of airless rocky exoplanets. The analogs of airless or nearly airless rocky exoplanets in the Solar System are the Moon, Mars, Mercury and asteroids, whose surface compositions have been studied extensively by spectroscopy of reflected solar radiation and planetary thermal emission (see Pieters \& Englert 1993 and de Pater \& Lissauer 2001 and extensive references therein). For investigations of the Solar System in 1970s and 1980s, the rocky bodies were still spatially unresolved or poorly resolved with resolution of a few thousand kilometers, and spectra of reflected solar radiation in the near-infrared (NIR) and spectra of planetary emission in the mid-infrared (MIR) were the only information to infer their surface composition. Thanks to prominent absorption features of specific minerals, notably olivine, pyroxene and ices, infrared spectroscopy provided significant constraints on the surface composition of rocky planets of the Solar System (e.g., Pieters \& Englert 1993).

We propose that using infrared spectroscopy to characterize mineral composition is applicable to study solid surfaces of exoplanets. Both reflected stellar radiation and planetary thermal emission from an exoplanet are potentially observable by secondary eclipses if the planet is transiting, as well as by direct imaging (e.g., Seager \& Sasselov 2000; Seager 2010). Still, characterization of exoplanetary surfaces is very different from the investigation of the Solar System analog surfaces, due to the fact that an exoplanet cannot be spatially resolved and the spectroscopy of exoplanets is limited to very low spectral resolutions, probably broad-band photometry in the near term. As a result, it is essential to focus on the most prominent spectral features in the disk-averaged radiation from an exoplanet.

We develop a theoretical framework to compute disk-integrated spectra of airless rocky exoplanets that self-consistently treats reflection and thermal emission and investigate the spectral features that can be used to interpret the mineral composition on exoplanetary surfaces. So far, the study of spectral features due to exoplanets' surfaces has been limited to features in the reflected light by vegetation (see Seager et al. 2005 for the ``red-edge") and a liquid water ocean (e.g., Ford et al. 2001; Cowan et al. 2009). In this paper we employ a generalized approach to investigate spectral features of solid materials on an exoplanet's surface, with a consistent treatment of reflected stellar irradiation and planetary thermal emission.

We focus on airless rocky exoplanets with solid surfaces, whose surface temperatures are lower than the melting temperature of silicates and other common minerals ($\sim1000$ - 2000 K). This surface temperature requirement ensures that the planetary thermal emission and the reflected stellar radiation can be separated in wavelengths. As shown in Figure \ref{SubStellar}, Kepler-22b, Kepler-20f and Kepler-11b can have an unmelted silicate surface. Moreover, a large number of Kepler planetary candidates have the orbital distances that permit unmelted silicate surfaces. In this paper we do not consider close-in rocky exoplanets that have molten lava surfaces, such as the case of Corot-7b. Once melted, crystal-field features, such as Fe$^{2+}$ electronic transition at 1 $\mu$m, will no longer persist in the same manner as in crystalline minerals. {\it In situ} measurements of active lava flow in Hawaii confirm that NIR spectra of molten lava are dominated by black-body emission (e.g., Flynn \& Mouginis-Mark 1992). We will address the spectral features of molten lava on close-in rocky exoplanet in a seperate paper.

The paper is organized as follows. In \S~2 we describe the background of geological surface remote sensing and the application in the study of the reflectance spectra of Solar System rocky planets. In \S~3 we define several types of planetary solid surfaces and describe a theoretical framework to compute the disk-average reflection and thermal emission spectra of an airless rocky exoplanet. \S~4 contains the results of the paper in which we present reflection and thermal emission spectra of exoplanets with various types of crust, and summarize the most prominent and diagnostic features. In \S~5 we discuss the effect of atmospheres, the effect of space weathering, detection potential of the proposed spectra, and the connection between the surface characterization and the planetary formation and evolution. We summarize the paper in \S~6.

\section{Background}

\subsection{Spectral Features of Geological Mineral Surfaces}

The spectral reflectances of minerals have features specific to their chemical compositions and crystal structures. From the visible to the NIR wavelengths (VNIR; 0.3 - 3 $\mu$m), minerals such as pyroxene, olivine and hematite create prominent absorption features in reflection data, mostly due to electronic transitions of transition element ions (i.e., Fe$^{2+}$, Fe$^{3+}$, Ti$^{3+}$, etc.) in the the crystal structures of minerals (Burns 1993). For example, Fe$^{2+}$ in the crystal field of olivine absorbs strongly at 1 $\mu$m. Position, strength, and number of features of the 1-$\mu$m absorption band are diagnostic of the relative proportions of the Mg and Fe cations in olivines (e.g., Sunshine \& Pieters 1998). In the ultraviolet (UV) and visible wavelengths, many minerals, such as hematite (Fe$_2$O$_3$), have strong charge-transfer absorption bands, making them very dark (e.g., Clark 1999). For secondary minerals, typically formed by interactions with water, their volatile components, such as hydroxyl (OH), water (H$_2$O), and carbonates (CO$_3^{2+}$), absorb strongly in the NIR due to overtone and combination absorptions from vibrations, i.e. bands and stretches of components in the crystal structure (e.g., Farmer 1974). See Table \ref{Mineral_List} for a list of common minerals on terrestrial planet surfaces and their key features in the NIR reflectance spectroscopy.

\begin{table}[htdp]
\scriptsize
\caption{Common minerals on terrestrial planet surfaces and their key features in the NIR reflectance spectroscopy. Spectra used are from representative samples of average grain diameters less than 200 $\mu$m, comparable to lunar soil (Duke et al., 1970).}
\begin{center}
\begin{tabular}{lll}
\hline
Mineral & Chemical Formula & Key Features \\
\hline
Olivine (Forsterite, 89Fo)\tablenotemark{a} 					& (Mg$_{0.89}$,Fe$_{0.11}$)$_2$SiO$_4$  & Absorption band at 1 $\mu$m  \\
Low-calcium pyroxene (Enstatite)\tablenotemark{b}  	& (Mg$_{0.8}$,Fe$_{0.2}$)SiO$_3$  & Absorption bands at 0.9 $\mu$m and 1.9 $\mu$m \\
High-calcium pyroxene (Augite)\tablenotemark{a}  		& (Ca,Na)(Mg,Fe,Al)(Si,Al)$_2$O$_6$  & Absorption bands at 1 $\mu$m and 2.3 $\mu$m \\
Plagioclase (Anorthite)\tablenotemark{a}  				& CaAl$_2$Si$_2$O$_8$ & High albedo, flat spectrum, absorption at 1.3 $\mu$m \\	
Hydrated silicate (Saponite)\tablenotemark{a} 		& Ca$_{0.25}$(Mg,Fe)$_3$(Si,Al)$_4$O$_{10}$OH$_2$$\cdot$n(H$_{2}$O)  & Absorption bands at 1.4 $\mu$m, 1.9 $\mu$m and 2.3 $\mu$m \\
Hematite\tablenotemark{a}  										& Fe$_2$O$_3$ & Low albedo, absorptions at 0.53 $\mu$m and 0.86 $\mu$m \\
Water ice\tablenotemark{c}   										& H$_2$O\tablenotemark{d}  & Absorption bands at 1.5 $\mu$m and 2 $\mu$m \\
\hline
\end{tabular}
\end{center}
\tablenotetext{a}{ Reflectance data are from the  USGS Digital Spectral Library (Clark et al. 2007).}
\tablenotetext{b}{ Reflectance data are from the RELAB Spectral Database (2010).}
\tablenotetext{c}{ Reflectance data are from Calvin \& Clark (1991). }
\tablenotetext{d}{ H$_2$O and OH also have strong fundamentals at about 3 $\mu$m (H$_2$O stretch, OH stretch, OH bend) and 6 $\mu$m (H$_2$O bend).}
\label{Mineral_List}
\normalsize
\end{table}

Similar to reflected stellar radiation, planetary thermal emission is encoded with information about the planet's surface composition. Absorption in the mid-infrared (MIR; 3 - 25 $\mu$m) is due to vibrational motions in crystal lattices, so that their wavelengths are related to the crystal structure and elemental composition (i.e., mineralogy) (Farmer 1974). Silicates have the most intense (i.e., the greatest absorption coefficient) spectral features between 8 and 12 $\mu$m due to the Si-O stretching, and the second most intense features between 15 and 25 $\mu$m due to the Si-O bending or deformation (Salisbury 1993). These vibration bands are so strong that they can manifest as mirror-like reflectance peaks and therefore emittance troughs. The shapes of these bands are complicated by the so-called ``transparency feature", features associated with a change from surface to volume scattering, and therefore are sensitive to particle size (Salisbury \& Walter 1989; Salisbury 1993). At the short-wavelength edge of the Si-O stretching band ($7.5\sim9.0$ $\mu$m), there is always a reflectance minimum (emittance maximum) as the refractive index of the mineral approaches that of the medium. This emittance maximum is termed ``Christiansen Feature" (CF), unique and ubiquitous for silicates. The wavelength of the CF is indicative of silica content of the material, i.e., more mafic silicates have the CF at longer wavelengths (Salisbury \& Walter 1989; Walter \& Salisbury 1989). For example, Glotch et al. (2010) inferred highly silica-rich compositions on the Moon by determining the CF wavelengths. Iron oxides also show spectral features due to Fe-O fundamentals, but at longer wavelengths than Si-O fundamentals, because iron is more massive than silicon (Clark 1999). For example, hematite, Fe$_2$O$_3$, has 3 strong stretching modes between 16 and 30 $\mu$m.

Spectral features of pure particulate minerals discussed above can be wide and deep, and could stand out in low-resolution and low signal-to-noise ratio spectra. In electronic processes, the electron participating in transitions may be shared between individual atoms and energy levels of shared electrons become smeared over wide energy bands, which results in wide spectral features (Burns 1993; Clark 1999). In vibrational processes, absorption bands are typically narrower than electronic features but can be broadened if the crystal is poorly ordered or if bands overlap (Farmer 1974; Clark 1999).

In reality, the contrast of spectral features may be significantly reduced, especially for thermal emission in MIR. Typical thermal emission band contrasts on planetary surfaces are less than $\sim$0.1 in the Solar System (Sprague 2000; Christensen et al. 2000; Clark et al. 2007). The reduction of spectral contrast in the thermal emission is mostly due to multiple scattering between regolith particles, with contributions of surface roughness and volume absorption (Hapke 1993; Kirkland et al. 2003). Multiple reflection, as a result of surface roughness, increases the emergent emissivity ($\epsilon_e$) towards unity, as illustrated by
\begin{equation}
\epsilon_e = 1 - (1-\epsilon)^{(n+1)} \ , \label{SurfRough}
\end{equation}
where $\epsilon$ is the true emissivity of material and $n$ is the number of reflections (Kirkland et al. 2003). As a result, unique identification of mineral compositions using planetary thermal emission usually requires high signal-to-noise ratio and high resolution spectroscopy.

Finally, mixing of minerals and space weathering may also reduce contrast of the spectral features. There are two levels of mixing, microscopic and macroscopic. Microscopic mixing (or intimate mixing) concerns particulates of different minerals that may be intimately mixed, such that photons are multiply-scattered by interactions with materials of different compositions. This is especially relevant for reflectance in VNIR. Macroscopic mixing (or areal mixing) concerns the planetary surface viewed as a disk average, which may contain discrete patches of different minerals or mineral assemblages. The average of macroscopic mixing can be modeled as a linear combination of reflected flux, whereas the microscopic mixing has to be treated as a multi-component radiative transfer problem. For spatially unresolved exoplanets, both microscopic mixing and macroscopic mixing need to be considered. Space weathering does not change the bulk mineralogic composition of the rocks and soils, but leads to formation of nanophase iron throughout the whole of the mature portion of the regolith, occurring in both vapor-deposited coatings on grain surfaces and in agglutinate particles, which alters the spectral properties of the surface significantly (e.g., Pieters et al. 2000; Hapke 2001).

\subsection{Solar System Airless Body Surface Spectra}

\begin{figure}[h]
\begin{center}
 \includegraphics[width=0.8\textwidth]{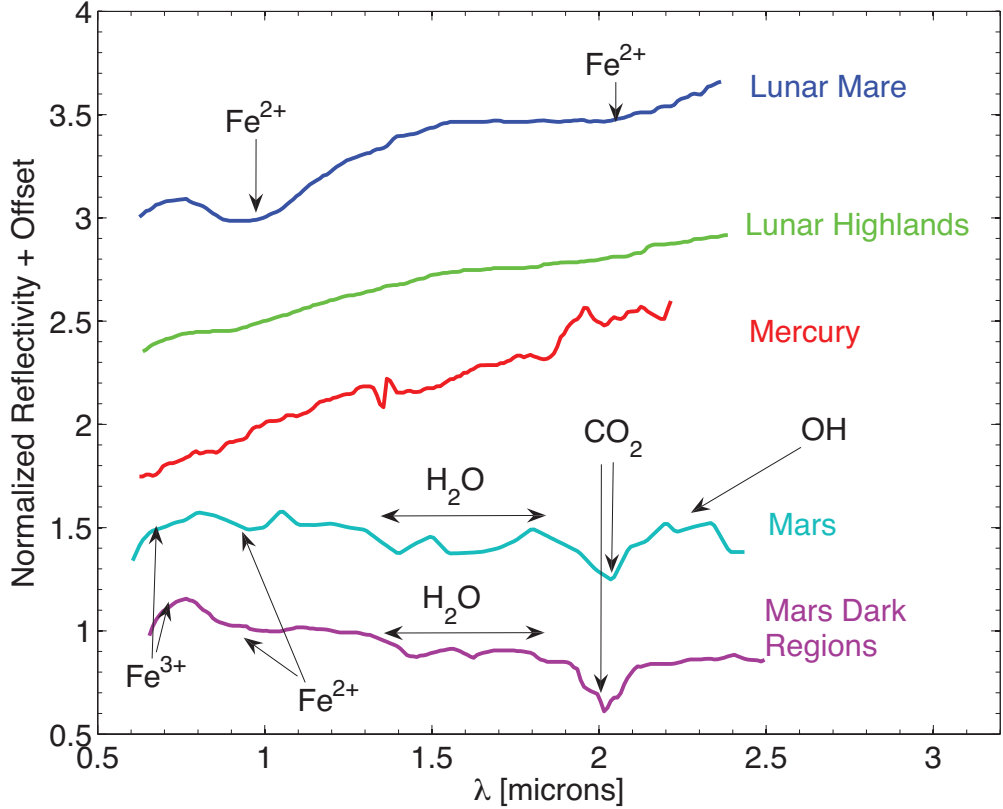}
 \caption{Ground-based telescopic reflectance spectra of the Moon, Mars and Mercury.
 The representative spectra shown on the figure are: (1) Lunar Mare Serenitatis, immature surface revealed by impact cratering, data from Pieters (1986); (2) Lunar Highland Descartes, immature surface revealed by impact cratering, data from Pieters (1986); (3) Mercury, disk average from bright limb to terminator, data from Blewett et al. (1997); (4) Mars, disk average, data from McCord \& Westphal (1971); (5) Mars, a dark region near equator of $\sim1000$ km in diameter, data from McCord et al. (1982).
 Reflectivity is normalized to be 1 at 1 $\mu$m and offset for clarity. Absorption of olivine and pyroxene (\ce{Fe^{2+}}) manifests in the lunar mare spectrum at 1 $\mu$m and 2 $\mu$m,
 and also shows in the Mars spectra at 1 $\mu$m. Mars' atmospheric \ce{CO2} imprints a deep absorption band at 2 $\mu$m. Mars spectra also contain strong charge-transfer absorption of \ce{Fe^{3+}} at wavelengths shorter than 0.75 $\mu$m, wide absorption feature from 1.4 to 1.7 $\mu$m, attributed to water ice on the surface, and a weak absorption near 2.2 $\mu$m, interpreted as hydrated minerals.
 }
 \label{SS}
  \end{center}
\end{figure}

Infrared spectral features have been used to study surface compositions of Solar System rocky planets. In the early-stage investigations, ground-based or balloon-based telescopic observations have been used to characterize the surfaces, for example the Moon (e.g., Murcray et al. 1970; Pieters 1978; McCord et al. 1981; Pieters 1986; Tyler et al. 1988), Mars (e.g., Singer et al. 1979; McCord et al. 1982) and Mercury (e.g., Vilas 1985; Tyler et al.1988; Blewett et al. 1997; Sprague et al. 2002). As a benchmark for the exoplanet investigations, we focus here on a description of the most prominent features in the low-spatial-resolution reflectance spectra and their implications. Several representative ground-based spectra of the Moon, Mars, and Mercury and the prominent features therein are shown in Figure \ref{SS}.

Reflectance spectra at NIR have been used to characterize the lunar mare and the lunar highlands. The lunar mare are generally dark and absorb strongly at 1 $\mu$m and 2 $\mu$m, which indicate they are of basaltic composition (e.g., Pieters 1978; see Figure \ref{SS}). The 1-$\mu$m absorption is due to iron-bearing glass, pyroxene, and olivine, and the 2-$\mu$m absorption is exclusively due to pyroxene (Pieters 1978). Moreover, the relative strength of the 1-$\mu$m absorption and the 2-$\mu$m absorption, as well as the band center positions of these absorptions, are used to infer the amount of olivine versus pyroxene and pyroxene composition of the lunar mare (McCord et al. 1981). The basaltic nature of the lunar mare indicates that  they formed by volcanic eruptions. In contrast to the lunar mare, the lunar highlands are bright and have nearly flat spectra at NIR with weak 1-$\mu$m and 2-$\mu$m absorptions, exhibiting a plagioclase composition with minor amounts of pyroxene (e.g., Pieters 1986; see Figure \ref{SS}). The nearly pure plagioclase composition of the lunar highlands indicates that they are primary crust formed from solidification of a magma ocean (e.g., Warren 1985).

Spatially unresolved spectroscopy of Mars has provided essential information to determine the surface composition of the red planet. Mars spectra feature strong Fe$^{3+}$ charge-transfer and crystal field absorptions from the near-UV to about 0.75 $\mu$m (McCord \& Westphal 1971; Singer et al. 1979; see Figure \ref{SS}). These spectral features, together with the visual red color and the polarization properties of Mars, established that the major component of the Martian surface is ferric oxides (e.g., hematite). Also, the spectra from dark areas on Mars show features of mixtures of pyroxene and olivine, probably covered by a layer of ferric oxides (McCord et al. 1982; see Figure \ref{SS}). The mineral composition of the Martian surface is interpreted as a result of secondary basaltic volcanism, affected by later oxidative weathering, perhaps in presence of liquid water to form Fe$^{3+}$ oxides (Mustard et al. 2005).

Telescopic spectra of Mercury suggest a surface similar to the lunar highlands. The NIR spectra of Mercury are flat without the signature Fe$^{+2}$ absorptions at 1 and 2 $\mu$m (Vilas 1985; see Figure \ref{SS}). Also, emission features in the mid-infrared (7 -- 13 $\mu$m) characteristic of silicate materials have been reported for Mercury (e.g., Tyler 1988). The spectroscopic evidence suggested a low-iron plagioclase surface similar to the lunar highlands (Blewett et al. 1997), consistent with the outcome of solidification of magma ocean without mantle overturn (Brown \& Elkins-Tanton 2009). Recent observations from a spacecraft orbiting Mercury have challenged this view by measuring Mg-rich, Fe-poor and Al-poor chemical compositions (ultramafic) and morphologies consistent with flood volcanism (Nittler et al. 2011; Head et al., 2011; Blewett et al. 2011). Research is on-going to explain Mercury data; a potential explanation reconciling all observations is atypically low iron content ($<\sim1$ \%) in common minerals forming ultramafic crusts (olivine, pyroxene), which usually exhibit Fe$^{+2}$ absorptions.

\section{Model}

\subsection{Types of Planetary Solid Surface}

The assemblage of minerals provides valuable information on the geological history and even the interior structure of the planet. For example, in the Solar System, feldspathic surfaces, such as the lunar highlands, are primitive products of crystallization from a magma ocean, since plagioclase is light in density and floats on top of the magma ocean (e.g., Warren 1985). For a relatively large planet with mass similar to that of Earth and Mars, the predicted crust composition after the overturn of the mantle in order to form a stable density stratification is dominated by Mg-rich olivines and pyroxenes, i.e., an ultramafic surface (Elkins-Tanton et al., 2005). Subsequent partial melting of mantle (i.e., volcanism) leads to production of distinctive igneous rocks such as basalts, i.e. the lunar mare. Finally, re-processing, heating and partial melting of these materials leads to the generation of granites, a regime on Earth driven by plate tectonics and incorporation of water in subducted crustal materials to lower the melting point (e.g., Taylor 1989).

We consider multiple geologically plausible planetary surface types as several assemblages of minerals, tabulated in Table \ref{Assemblage}. The types include primary crust, i.e., the crust that forms from solidification of magma ocean; secondary crust, i.e., the crust that forms from volcanic eruptions; and tertiary crust, i.e., the crust that forms from tectonic re-processing. Each type of igneous crust differs from another by assemblages of minerals, which are in turn governed by thermodynamics, planetary composition, and planetary history (e.g. Best 2002; Hazen et al. 2008). Moreover, various modification processes in the planetary evolution may alter the surface spectral properties significantly. The modification processes include aqueous alteration and oxidative weathering \footnote{Space weathering is discussed in section \ref{SpaceWeathering}}. Additionally we simulate an ice-rich planet and one lacking a silicate crust or mantle. The list in Table \ref{Assemblage} encompasses the most common surface types; abiotic mechanisms of surface formation and modification are considered, and the list largely covers the diversified solid surfaces on Solar System rocky planets.

\begin{table}[htdp]
\scriptsize
\caption{Potential crustal compositions of rocky exoplanets. NIR spectra were defined for eight notional exoplanet surface types from laboratory measurement of rock powders (sample) or radiative transfer modeling combining endmember mineral samples measured in the laboratory (modeled). }
\begin{center}
\begin{tabular}{lllll}
\hline
\hline
Type			& Mineral Composition 			& Spectrum & Indication & Solar System  \\
		&  			& Source &  & Examples \\
\hline
\hline
\multirow{2}{*}{Metal-rich} 		& \multirow{2}{*}{Pyrite}		& \multirow{2}{*}{Sample\tablenotemark{a}} 	& Primary crust & \multirow{2}{*}{N/A}\\
 		& 		&  	& with mantle ripped off & \\
 \hline
\multirow{3}{*}{Ultramafic}	& \multirow{3}{*}{60\% Olivine, 40\% Enstatite} & \multirow{3}{*}{Modeled\tablenotemark{b}}	 & Primary crust  & Primary Earth and Mars; \\
	& &	& with mantle overturn; &  and early Earth lavas \\
& & & \multicolumn{2}{l}{or secondary crust from hot lavas} \\
\hline
\multirow{2}{*}{Feldspathic }	& \multirow{2}{*}{97\% Fe-plagioclase, 3\% Augite}			& \multirow{2}{*}{Sample\tablenotemark{c}}	& Primary crust  & \multirow{2}{*}{Lunar Highlands}\\
 	& 		& & \multicolumn{2}{l}{without mantle overturn }  \\
 \hline
\multirow{3}{*}{Basaltic}	 	& 76\% Plagioclase, 8\% Augite,					& \multirow{3}{*}{Sample\tablenotemark{d}}	& \multirow{3}{*}{Secondary crust} & Lunar Mare and  \\
	 	& 6\% Enstatite, 5\% Glass,					& 	&  & locations on current Earth \\
 & 1\% Olivine & & &\\
\hline
\multirow{2}{*}{Granitoid}	 	& 40\% K-feldspar, 35\% Quartz					& \multirow{2}{*}{Modeled\tablenotemark{b}}	& \multirow{2}{*}{Tertiary crust}  & \multirow{2}{*}{Current Earth} \\
	& 20\% Plagioclase, 5\% Biotite					& &   &  \\
\hline
Clay	& 50\% Mg-smectite, 50\% Serpentine 	& Modeled\tablenotemark{b}	& Aqueously altered crust & Locations on current Earth and Mars \\
\hline
Ice-rich silicate			& 50\% Water ice, 50\% Basalt 	& Modeled\tablenotemark{b}	& Ice-rich silicate mantle & Locations on current Earth and Mars \\
\hline
Fe-oxidized		& 50\% Nanophase hematite, 50\% Basalt				& Modeled\tablenotemark{b}	& Oxidative weathering & Current Mars \\
\hline
\hline
\end{tabular}
\end{center}
\tablenotetext{a}{\ Clark et al. (2007). }
\tablenotetext{b}{\ Spectra are synthesized from the measured spectra of each endmember minerals in Clark et al. (2007).}
\tablenotetext{c}{\ Lunar anorthosite, sample 15415, Cheek et al. (2009).}
\tablenotetext{d}{\ Basalt sample 79-3b from Wyatt et al. (2001). }
\label{Assemblage}
\normalsize
\end{table}

As is evident from Table \ref{Assemblage}, a planet's surface is defined as the assemblage of several endmember minerals. Moreover, a planet's surface may be composed of bulk patches of different crusts, as is the case for the lunar mare and highlands. It is therefore essential to consider the macroscopic mixture of different crusts as well as the microscopic (intimate) mixture of different minerals for each crust.

We model the intimate mixture of minerals as follows. For each endmember mineral, we use the measured bidirectional reflectance in the USGS Digital Spectral Library (Clark et al. 2007) or the RELAB Spectral Database (2010). We retrieve the wavelength-dependent single scattering albedo ($\omega$) of each type of mineral from the experimental data based on an analytical radiative transfer model of Hapke (1981, 2002), which will be detailed in section 3.2. We average the mineral composition utilizing single scattering albedo spectra of endmembers, weighted by their mixing ratios. The bidirectional, directional-hemispherical reflectance and directional emissivity of the mixture can thus be computed using the Hapke radiative transfer model in the forward sense (Hapke 1981, 2002). This method of computing reflectance spectra of mineral mixtures has been proven within 10\% error by experiments for binary and ternary mixing among components with moderate albedo contrast (Mustard \& Pieters 1989).

There are some fundamental limitations of this approach to synthesize reflectance spectra of mineral mixtures. First, the Hapke radiative-transfer method is designed for mixture of particulate materials whose grain sizes are small and whose phase dependent scattering behaviors are similar. Although regoliths are widespread on the surfaces of rocky bodies in the Solar System as a result of extensive meteoritic bombardments, it is uncertain whether an exoplanet's surface is made of regolith or bulk rocks \footnote{We focus on regolith surfaces in this paper. Surfaces of an airless planetary body are most likely comprised of particulate regoliths, because impact gardening effectively converts surface rocks to a regolith layer.}. Second, the Hapke radiative-transfer method may induce an error up to 25\% for for dark materials (basalts) in mixtures with clay (Ehlmann \& Mustard, in preparation). For these reasons, we choose well-characterized actual materials rather than Hapke modeled spectra as representative of feldspathic and basaltic surface types (see Table \ref{Assemblage}), and results from intimate mixing studies should be considered indicative rather than exact.

\subsection{Bidirectional Reflectance Spectra of Planetary Surface Material}

\begin{figure}[h]
\begin{center}
 \includegraphics[width=0.4\textwidth]{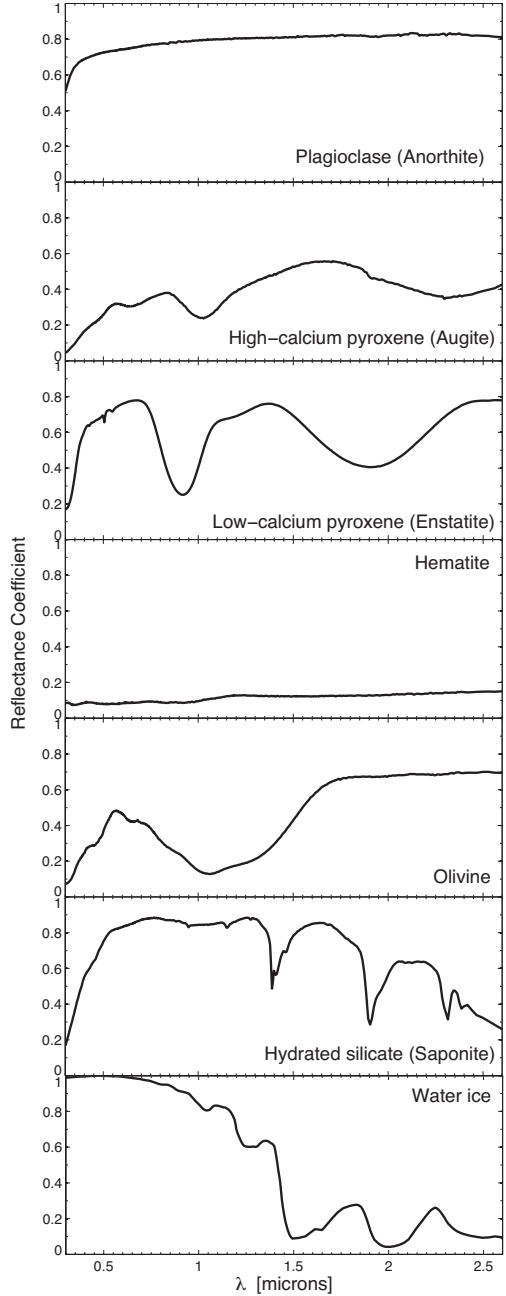}
 \caption{Radiance coefficient of common minerals on terrestrial planet surfaces listed in Table \ref{Mineral_List}.
 Bidirectional reflectance spectra measured at incidence angle $i=30^{\circ}$, scattering angle $e=0^{\circ}$ and phase angle $g=30^{\circ}$ (Clark et al. 2007; RELAB 2010; Calvin \& Clark 1991) are used to derive the radiance coefficient.
 Plagioclase (non-iron-bearing) has a bright and flat NIR spectrum and hematite has a dark and flat NIR spectrum.
 Mafic minerals, i.e., olivine and pyroxene, have a deep and wide electronic absorption at 1 $\mu$m. In addition, pyroxene
 has an absorption near 2 $\mu$m, and the band center location depends on its calcium content.
 Water ice has deep and wide vibrational absorption bands at 1.5 $\mu$m and 2 $\mu$m, whereas hydrated minerals have narrow vibrational absorption bands at 1.4, 1.9 and 2.3 $\mu$m.}
 \label{rc}
  \end{center}
\end{figure}

For computation of planetary surface spectra the most important parameter is bidirectional reflectance of surface solid material, which is usually characterized by the radiance coefficient ($r_c$). The radiance coefficient is defined as the brightness of a surface relative to the brightness of a Lambert surface identically illuminated, which depends on the direction of both incident and scattered light. The radiance coefficient of a solid material not only depends on its chemical composition, but also depends on its crystal structures. Reflectance spectra of these minerals are shown in Figure \ref{rc}.

Hapke (1981, 2002) presents a straightforward method to compute the approximated radiance coefficient of any particulate material in terms its single scattering albedo. In essence, Hapke (1981, 2002) treats the problem as the radiative transfer of a planar, semi-infinite, particulate medium illuminated by collimated light. The key assumption to achieve a convenient analytical expression is to assume isotropic scatterers for multiple scattering and non-isotropic scatterers for single scattering (Hapke 1981). The resulting expression has been proved to be correct and handy in the interpretation of surface composition of the Moon and Mars (e.g., Hapke 2002).

Experimental data of radiance coefficient ($r_c$) of common minerals and mixtures are typically presented for certain combinations of incidence angle ($\mu_0$), scattering angle ($\mu$) and phase angle ($g$). For application to exoplanets, we extend the experimentally measured  $r_c$ to any combination of $(\mu_0,\mu,g)$ using the analytical expression of Hapke (2002). Also, we compute the directional-hemispherical reflectance ($r_{\rm h}$) for different minerals based on experimental bidirectional reflection using an analytical expression given by Hapke (2002). The analytical form of Hapke (2002) employs a parameter $h$ to describe the opposition effect and a phase function to describe the single particle scattering. For fine grains, $h\sim0.1-0.4$. We assume $h=0.2$ and the phase function to be the first order Legendre expansion as
\begin{equation}
p(g) = 1+b\cos(g) \ ,
\end{equation}
where $b$ is the anisotropy parameter in $[-1,1]$. We typically assume $b=0$, i.e. isotropic scattering. Numerical experiments show that the retrieval of single scattering albedo and the computation of radiance coefficient do not sensitively depend on the assumptions of $h$ and $b$.

For consistent treatment of reflection and thermal emission, we use experimentally measured reflectance and derive emissivities of mineral assemblages. In the Hapke framework, the directional emissivity and the directional-hemispherical reflectance obey Kirchhoff's Law (Hapke 1993). We performed cross-checks between our derived emissivity and experimentally measured emissivity of pure minerals (Christensen et al. 2000) and demonstrate that they are roughly consistent.

\subsection{Disk-Averaged Spectral Model}

Where we differ from the solar-system models are that spectra of exoplanets are always disk integrated. The observable quantity of exoplanet reflection is the occultation depth of the secondary eclipse. Reflected light from an exoplanet is an average over the entire sub-stellar hemisphere. Moreover, the planetary radiation flux combines the reflected stellar radiation and the thermal emission, as these two components may overlap in wavelength ranges. Here we present a detailed formulation of disk-averaged reflection and thermal emission spectra from an airless rocky exoplanet with contribution from thermal emission, in terms of the radiance coefficient ($r_c$) and the directional-hemispherical reflectance ($r_{\rm h}$) of its surface materials.

Radiative flux per unit area per wavelength from an exoplanet for an Earth-based observer $F_{\rm p}$ (erg s$^{-1}$ cm$^{-2}$ nm$^{-1}$) is a hemisphere-integral as
\begin{equation}
F_{\rm p} = \bigg(\frac{R_{\rm p}}{D}\bigg)^2\int_{-\frac{\pi}{2}}^{\frac{\pi}{2}}\int_{-\frac{\pi}{2}}^{\frac{\pi}{2}}
I_{\rm p}(\theta,\phi)\cos^2\theta\cos\phi\ d\theta d\phi\ ,
\end{equation}
in which $R_{\rm p}$ is the radius of the planet, $D$ is the distance to the observer, and $I_{\rm p}(\theta,\phi)$ is the intensity from a location on the hemisphere towards the observer specified by latitude-longitude coordinates $(\theta,\phi)$. The coordinate system is chosen such that the observer is at the direction of $(\theta=0,\phi=0)$. The planetary radiance is composed of the reflection of stellar light and the thermal emission from the planet itself,
\begin{equation}
I_{\rm p}(\theta, \phi) = I_{\rm s}(\theta, \phi) + I_{\rm t}(\theta, \phi) \ .\label{combi}
\end{equation}

The reflected intensity is related to the incident stellar irradiance ($F_{\rm inc}$) as
\begin{equation}
I_{\rm s}(\theta, \phi) = F_{\rm inc}\frac{\mu_0}{\pi} r_c(\mu_0,\mu,g) \ ,
\end{equation}
where $r_c$ is the radiance coefficient as a function of the incidence angle $i$, the scattering angle $e$ and the phase angle of scattering $g$. By definition, a Lambertian sphere has $r_c=1$. Let $\alpha$ be the phase angle of the exoplanet with respect to the Earth, so that the stellar coplanar light comes from the direction of $(\theta=0,\phi=\alpha)$. For each surface element, we have the following geometric relations :
\begin{eqnarray}
&& \mu_0\equiv\cos i = \cos\theta\cos(\alpha-\phi) \ ,\\
&& \mu  \equiv\cos e = \cos\theta\cos\phi \ ,\\
&& g = \alpha \ .
\end{eqnarray} The thermal emission of the planet depends on its surface temperature, which is controlled by both stellar radiation and planetary surface properties. The irradiance incident on the planet is
\begin{equation}
F_{\rm inc} = \pi B_{\lambda}[T_*]\bigg(\frac{R_*}{D_{\rm p}}\bigg)^2\ ,
\end{equation}
and the stellar irradiance for the observer is
\begin{equation}
F_{*} = \pi B_{\lambda}[T_*]\bigg(\frac{R_*}{D}\bigg)^2\ ,
\end{equation}
in which $T_*$ and $R_*$ are the temperature and the radius of the star, $D_{\rm p}$ is the semi-major axis of the planet's orbit, and $B_{\lambda}$ is the Planck function for blackbody radiance.
The thermal emission intensity is
\begin{equation}
I_{\rm t} (\theta, \phi) = \epsilon_{\lambda}(\mu)B_{\lambda}[T(\theta, \phi)]\ ,
\end{equation}
where $\epsilon_{\lambda}$ is the directional emissivity, and $T(\theta, \phi)$ is the temperature of the planet's surface. According to the Kirchhoff's law of thermal radiation and the framework of Hapke (1993), $\epsilon_{\lambda}$ is tied to the directional-hemispherical reflectance $r_{\rm dh} $ via
\begin{equation}
\epsilon_{\lambda} (\mu) = 1 - r_{\rm dh}  (\mu) \ .
\end{equation}
An effectively airless rocky exoplanet, without efficient heat transport mechanisms, is likely to have local thermal equilibrium (e.g. L\'eger et al. 2011 for CoRot-7b). The energy balance equation can therefore be written as
\begin{equation}
\mu_0 \int \epsilon_{\lambda} (\mu_0) F_{\rm inc} d\lambda = \pi \int \epsilon_{\lambda}^{h} B_{\lambda}[T] d\lambda \ , \label{LocalBalance}
\end{equation}
where $\epsilon_{\lambda}^{h}$ is the hemispheric emissivity, i.e., the hemispherical average of the directional emissivity. By solving this equation we determine the local surface temperature and then compute the directional thermal emission of each surface element.

Finally, the occultation depth of the secondary eclipse is
\begin{equation}
\frac{F_{\rm p}}{F_*} = \frac{1}{F_{\rm inc}} \int_{-\frac{\pi}{2}}^{\frac{\pi}{2}}\int_{-\frac{\pi}{2}}^{\frac{\pi}{2}}
I_{\rm p}(\theta,\phi)\cos^2\theta\cos\phi\ d\theta d\phi\ \times \bigg(\frac{R_{\rm p}}{D_{\rm p}}\bigg)^2 \equiv A_{\rm g} \bigg(\frac{R_{\rm p}}{D_{\rm p}}\bigg)^2 \ , \label{AG}
\end{equation}
where $A_{\rm g}$ is the apparent geometric albedo of the planet, and $I_{\rm p}(\theta,\phi)$ should be evaluated from Equation (\ref{combi}) at $\alpha=0$. Here we include thermal emission into the definition of geometric albedo, because for close-in exoplanets the reflected stellar light and the thermal emission may not be separated in spectra. In case of negligible thermal emission, for example in the visible and NIR wavelengths for Earth-like planets, the geometric albedo in Equation (\ref{AG}) can be simplified to be the conventional definition as
\begin{equation}
A_{\rm g} = \frac{1}{\pi}\int_{-\frac{\pi}{2}}^{\frac{\pi}{2}}\int_{-\frac{\pi}{2}}^{\frac{\pi}{2}}
r_c(\mu_0,\mu_0,0)\cos^3\theta\cos^2\phi\ d\theta d\phi\ \ ,
\end{equation}
consistent with Sobolev (1972) and Seager (2010). In addition, we use the apparent brightness temperature ($T_{\rm b}$) to describe the thermal emissivity of the planet, namely
\begin{equation}
\pi B_{\lambda}[T_{\rm b}] = F_{\rm p} \ .
\end{equation}
The brightness temperature defined as such can be compared with observations directly, and takes into account the disk average of directional emissivities and surface temperatures.

\section{Results}

Our main findings are: rocky silicate surfaces lead to unique features in planetary thermal emission at the mid-infrared due to strong Si-O vibrational bands (7 - 13 $\mu$m and 15 - 25 $\mu$m); the location of the emissivity maxima at the short-wavelength edge of the silicate feature (7 - 9 $\mu$m) is indicative of the silica content in the surface silicates; ultramafic surfaces can be uniquely identified in the reflectance spectra via a prominent absorption feature at 1 $\mu$m (i.e., the J band); hydrous surfaces induce strong absorption at 2 $\mu$m (i.e., the K band); and surface water ice has a unique absorption feature in the reflected stellar light at 1.5 $\mu$m (i.e., the H band).

\begin{figure}[h]
\begin{center}
 \includegraphics[width=1 \textwidth]{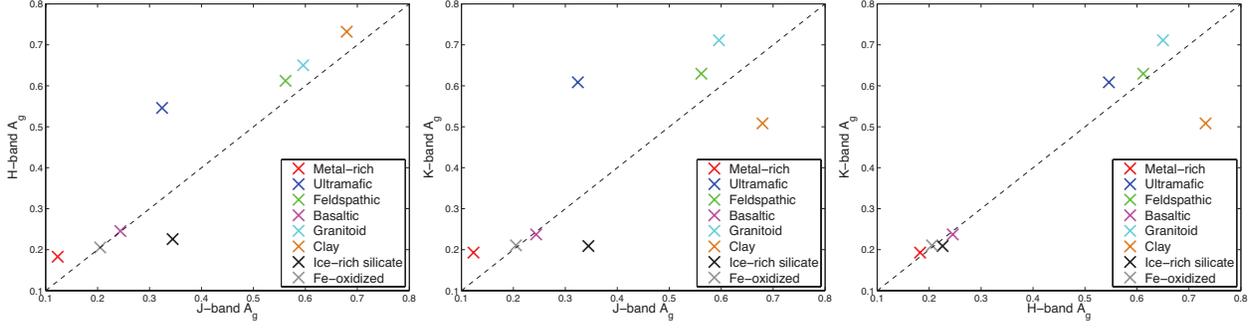}
 \caption{Comparison between geometric albedos in J, H, and K bands for airless exoplanets with 8 types of crust listed in Table \ref{Assemblage}. The dashed line indicates a flat spectrum. The thermal emission is not included.
 Note that ultramafic surfaces have J-band albedo significantly smaller than H-band and K-band albedo, ice-rich silicate surfaces have H-band and K-band albedo significantly smaller than J-band albedo, and clay surfaces have K-band albedo significantly smaller than J-band and H-band albedo.}
 \label{crust_scatter}
  \end{center}
\end{figure}

In the following we present disk-average spectra of airless rocky exoplanets and describe the main results. We compute the disk average using Equation (\ref{AG}) based on measured or modeled bidirectional reflectance of mineral assemblages as tabulated in Table \ref{Assemblage}. Figure \ref{FullCase} shows the VNIR geometric albedos and MIR brightness temperatures of airless exoplanet fully covered by the 8 types of crust; Table \ref{TIR_Features} lists the main spectral features due to surface minerals in the planetary thermal emission; Table \ref{crust_bb} lists the geometric albedos of the 8 cases averaged in the NIR J (1.1 - 1.4 $\mu$m), H (1.5 - 1.8 $\mu$m), and K (2 - 2.4 $\mu$m) bands; Figure \ref{crust_scatter} is a scatter plot showing the relation between broad-band true geometric albedos for the 8 cases; Figure \ref{crustmix} explores the effect of macroscopic mixture of two types of crust; and Figure \ref{Kepler20f} shows the modeled planetary spectra of Kepler-20 f if the planet has a particulate solid surface.

\subsection{Silicate Features in the Thermal Emission of Rocky Exoplanets}

\begin{figure}[h]
\begin{center}
 \includegraphics[width=1 \textwidth]{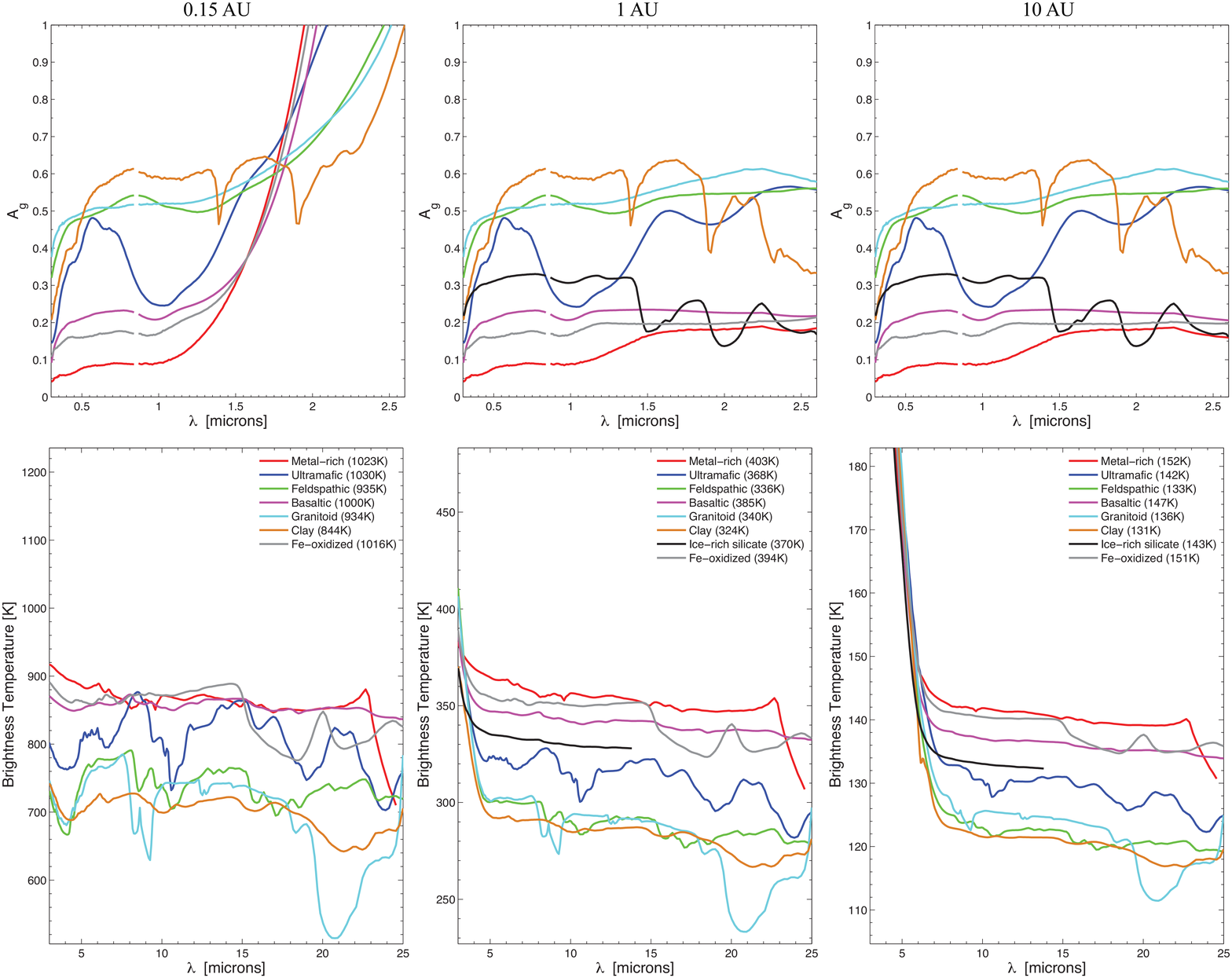}
 \caption{
 Apparent geometric albedo in VNIR and brightness temperature in MIR of an airless exoplanet fully covered by the 8 types of crust listed in Table \ref{Assemblage}. The three planetary scenarios correspond to solar-like host star and semi-major axis $D_{\rm p} = 0.15$ AU, 1 AU and 10 AU. For a semi-major axis of 1 and 10 AU, the planetary flux in VNIR is purely due to reflection and the geometric albedo is the true $A_{\rm g}$ independent of stellar irradiance; for close-in exoplanets (e.g., $D_{\rm p} = 0.15$ AU), thermal emission extends to about 1 $\mu$m. In parentheses we list the sub-stellar temperatures on the planet's surface.
}
 \label{FullCase}
  \end{center}
\end{figure}

\begin{figure}[h]
\begin{center}
 \includegraphics[width=0.8\textwidth]{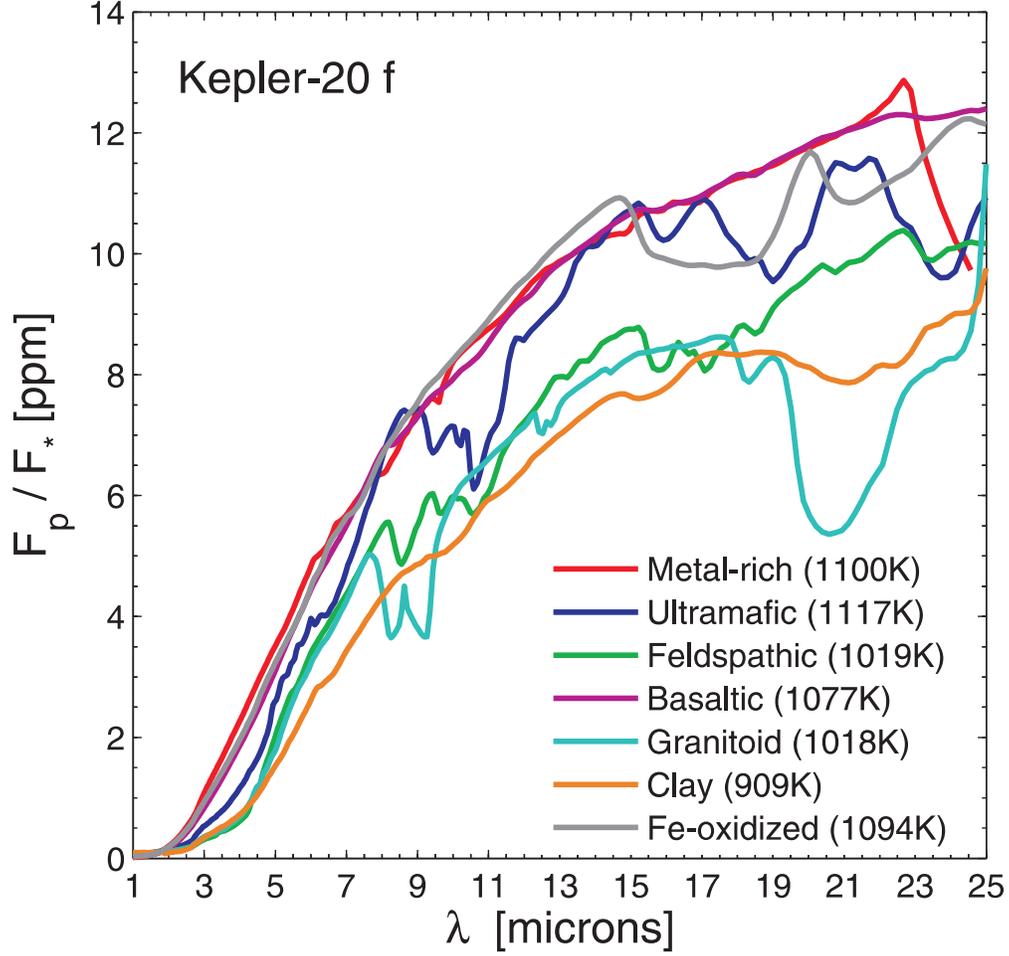}
 \caption{
Modeled transit depth of the secondary eclipse of Kepler-20f, if the planet's surface is covered by particulate materials as listed in Table \ref{Assemblage}. We use the planetary parameters of Fressin et al. (2012) and find that in the Kepler's bandpass the secondary transit depth is less than 1 part-per-millon (ppm). Sub-stellar temperatures listed in the figure are self-consistently computed in the model, and we verify that except for the granitoid and clay the surfaces are solid anywhere on the planet. We note that the Si-O features lead to variations in secondary transit depth as large as 2 ppm in both the 7 - 13 $\mu$m band and the 15 - 25 $\mu$m band, for ultramafic and granitoid surfaces. The iron-oxide feature in the 15 - 25 $\mu$m band and the pyrite feature at $>22$ $\mu$m are also evident.
 }
 \label{Kepler20f}
  \end{center}
\end{figure}

Silicate surfaces possess prominent minima in the thermal emission spectra from 7 - 13 $\mu$m and 15 - 25 $\mu$m (see Figure \ref{FullCase}). These Si-O stretching and bending vibrations manifest in the thermal emission spectra as troughs of complicated shapes, due to a strong reststrahlen reflection at the band center and volume scattering near the band edges. Prominent Si-O features, e.g. in the ultramafic and granitoid surfaces, have equivalent width (EW) larger than 1 $\mu$m and $\Delta T_{\rm b}$ larger than 20 K (see Table \ref{TIR_Features}). For comparison, the atmospheric O$_3$ absorption line at 9.6 $\mu$m has $\Delta T_{\rm b}$ of about 30 K for the Earth (Des Marais et al. 2002; Belu et al. 2011). For close-in rocky exoplanets, $\Delta T_{\rm b}$ of the silicate features can be as large as of 200 K (see Figure \ref{FullCase}), which corresponds to a variation of secondary transit depth of 2 part-per-million (ppm) for the case of Kepler-20f (see Figure \ref{Kepler20f}). In contrast to silicate surfaces, iron-oxidized surfaces do not have thermal emission troughs in 7 - 13 $\mu$m, but usually have a clear double-peak Fe-O feature in the 15 - 25 $\mu$m band, as shown in Figure \ref{FullCase} for the Fe-oxidized surface. Silicates and iron-oxides can therefore be distinguished based on thermal emission spectra. In summary, wide troughs of brightness temperature in both 7 - 13 $\mu$m and 15 - 25 $\mu$m constitute a unique signature of silicate surfaces of exoplanets.

Furthermore, high-resolution spectra of the silicate bands can allow identification of different kinds of silicate surfaces. As shown in Table \ref{TIR_Features}, as silica content in the mineral assemblage increases, the Christiansen feature (CF), defined as the emissivity maxima at the edge of the main Si-O band, shifts to shorter wavelengths. Note that the ultramafic surface is the most silica-poor and the granitoid surface is the most silica-rich. This well-known effect in mineralogy is applicable to spectral analysis of disk-integrated planetary thermal emission for characterizing rocky surfaces. Determination of the CF wavelength, however, involves tracing the curvature of brightness temperature spectra and therefore requires high-resolution spectra with a high signal-to-noise ratio. At a minimum, the determination of CF location and thus silica content requires three narrow bands in the 7 - 9 $\mu$m region (see Greenhagen et al. 2010 and Glotch et al. 2000 for an example applicable to the Moon).

Not only the spectral features, but also the thermal emission continuum provide valuable information on the planetary surfaces. Thermal emission of an effectively airless rocky exoplanet probes the planet's surface temperature, which in turn depends on its surface composition. A highly reflective surface at VNIR, for example the feldspathic surface, has much lower equilibrium  temperature than a VNIR absorptive surface, for example the basaltic surface (e.g. Figure \ref{FullCase}). For an airless Earth analog, the brightness temperature continuum from 5 to 25 $\mu$m is about 340 K if the surface is basaltic and 290 K if the surface is feldspathic (see Figure \ref{FullCase}). Solely due to different VNIR reflectivities, the apparent temperature difference of the planet, $\Delta T_{\rm b}$, can be as large as 50 K. As a result, equilibrium temperature of planetary solid surfaces, derived from the planetary thermal emission, may coarsely constrain the planetary surface composition, for example to the level of feldspathic versus ultramafic. Note in Figure \ref{FullCase} that more than one types of crust have overall low VNIR albedos, including metal-rich, basaltic and Fe-oxidized crusts; several crusts have overall high VNIR albedos, including feldsphathic, granitoid and clay crusts; and several crusts have intermediate VNIR albedos, including ultramafic and ice-rich silicate crusts. Moreover, the overall surface reflectivity sensitively depends on the surface roughness and particle size. Intrinsic degeneracy between surface composition and roughness or particle size exists if thermal emission continuum, or equilibrium temperature, is the only piece of information. This highlights the utility of multiple spectral channels in characterizing exoplanet surfaces.

Last but not the least, we comment that thermal emission does not cause the reduction of spectral feature contrasts at NIR wavelengths, unless the exoplanet is very hot when in a close-in orbit (i.e., semi-major axis less than $\sim$0.2 AU). For an analog of Earth or Mars, the thermal emission is negligible at NIR. For close-in exoplanets, the sub-stellar temperature may be very high and their thermal emission may extend to the NIR if the planet's orbit is at 0.15 AU (e.g. Figure \ref{FullCase} upper left). In this case, the thermal emission may severely reduce the spectral features at $>2$ $\mu$m, because the thermal emission compensates for the absorption of stellar radiation. According to Kirchhoff's law, a surface has high emissivity at wavelengths where it absorbs strongly. For example, the strong absorption at 2.3 $\mu$m of hydrated silicates is largely compensated by the strong thermal emission at the same wavelengths (see Figure \ref{FullCase}). For even shorter orbital periods, the sub-solar point may reach the melting temperature of minerals on the surface, which creates molten lava on the surface (see Figure \ref{SubStellar}). For solar-type stars, the transition of solid surface and molten lava happens at about 0.1 AU. To date exoplanets that could be rocky mostly lie in the regime of molten lava. Notably, Kepler-20f is marginally at the melting point for a granitoid surface, and is certainly solid for more refractory surfaces such as ultramafic. Since it is unlikely that Kepler-20f possesses a significant gas envelope (Fressin et al. 2012), Kepler-20f may be a good exoplanetary candidate for solid surface characterization via prominent silicate and iron-oxide features (see Figure \ref{Kepler20f}).

\begin{table}[htdp]
\scriptsize
\caption{Spectral features in planetary thermal emission that are characteristic of surface compositions.}
\begin{center}
\begin{tabular}{l|l|lllll}
\hline
\hline
Crust Type & CF\tablenotemark{a} & Feature & Range\tablenotemark{c} & Emissivity Minima & EW\tablenotemark{d} & $\Delta T_{\rm b}$\tablenotemark{e}\\
& ($\mu$m) & & ($\mu$m) & ($\mu$m) & ($\mu$m) & (K) \\
\hline
Metal-rich	& NA & Pyrite band & $>22.7$  & -  & 0.27 & $>47.07$ \\
\hline
Ultramafic	& 8.5 & Si-O stretch & 8.5 - 13.7 & 10.6 & 0.60 & 25.0 \\
           			&        & Si-O bending\tablenotemark{b} & $>15.1$ & 16.1, 19.0, 23.9 & 1.38 & 8.9, 26.9, 40.6 \\
\hline
Feldspathic & 8.0 & Si-O stretch & 8.0 - 12.9 & 8.6, 10.6 & 0.47 & 9.4, 12.9 \\
           			&        & Si-O bending\tablenotemark{b} & $>15.4$ & 15.9, 17.1, 18.5, 23.5  & 0.67 & 8.0, 12.7, 10.9, 10.6 \\
\hline
Basaltic		& 8.0 & Si-O stretch & 8.0 - 12.9 & 10.8 & 0.13 & 3.9 \\
           			&        & Si-O bending\tablenotemark{b} & $>15.0$ & 15.9, 17.1, 18.5  & 0.35 & 2.9, 4.6, 5.9 \\
\hline
Granitoid	& 7.6 & Si-O stretch & 7.6 - 10.2 & 8.3, 9.3 & 0.48 & 18.3, 24.0 \\
           			&        & Si-O bending\tablenotemark{b} & $>14.8$ & 18.3, 20.9, 24.1  & 2.10 & 17.0, 57.1, 29.4 	 \\
\hline
Clay			& 8.0 & Si-O stretch & 8.0 - 13.0 & 10.1, 11.5 & 0.17 & 4.6, 2.6 \\
           			&        & Si-O bending\tablenotemark{b} & $>14.1$ & 15.8, 21.3, 22.5, 24.6  & 1.02 & 4.8, 20.5, 20.3, 14.1 \\
\hline
Ice-rich silicate& 7.9 & Si-O stretch & 7.9 - 12.8 & 10.9 & 0.05 & 1.3 \\
\hline
Fe-oxidized	& 8.1 & Si-O stretch & 8.1 - 11.9 & 10.8 &0.02 & 1.2 \\
           			&        & Fe-O stretch    & $>14.3$ & 18.5, 21.3  & 1.38 & 28.2, 25.5 \\
\hline
\hline
\end{tabular}
\tablenotetext{a}{The wavelengths of the Christiansen Feature are defined as the emission maxima at the short-wavelength edge of Si-O stretch fundamentals.}
\tablenotetext{b}{These complex spectral features are mainly due to Si-O bending and deformation, with minor contribution of Si-O-Si stretch.}
\tablenotetext{c}{For certain spectral features the wavelength ranges are not well defined, due to the complex nature of these features and limitations of experimental measurements at mid and far infrared wavelengths.}
\tablenotetext{d}{Equivalent widths of spectral features are defined as ${\rm EW}=\int(1-f/f_0)d\lambda$ where $f$ is the observable planetary flux and $f_0$ is the continuum.  The continuum is derived from linear interpolation of brightness temperature between the edges of features. When only one edge can be defined, the continuum is derived from the constant brightness temperature, and the actual equivalent width of the feature might be larger than tabulated.}
\tablenotetext{e}{$\Delta T_{\rm b}$ is the decrease of brightness temperature attributed to the spectral features with respect to the continuum, reported for each corresponding emissivity minima.}
\end{center}
\label{TIR_Features}
\end{table}

\subsection{General Reflection Spectra of Exoplanets with Solid Surfaces}

Planetary surfaces of different compositions can have very different overall reflectivity. As shown in Figure \ref{FullCase}, a pure feldspathic surface (e.g., the lunar highlands) is very bright, with a geometric albedo $\sim$0.6 and a relatively flat NIR spectrum. In contrast, basaltic, Fe-oxidized or metal-rich surfaces are very dark, with a geometric albedo less than 0.3. As shown in Figure \ref{crust_scatter}, when these spectra are downsampled to the resolution of the J, H, and K bands, there is a strong correlation between reflectivity in the three bands for most surface types.

The absorption band near 1 $\mu$m is deep and wide for the ultramafic surface. The constituents of this surface are olivine and pyroxene, both of which have Fe$^{2+}$ and absorb strongly near 1.0 $\mu$m. Weaker absorptions in this band can also be found in the feldspathic, basaltic, Fe-oxidized, and ice-rich silicate cases because they all contain some olivine or pyroxene in their composition. The absorptions are caused by ferrous iron, commonly present at $>1$\% levels in the igneous minerals comprising these surfaces (although see section 2.2 discussion on Mercury for the possibility of non-ferrous components). For the feldspathic crust, the absorption peak is shifted to $\sim$1.3 $\mu$m, which is distinctive for Fe-anorthite, a feldspar that has been positively identified in the lunar highlands (e.g., Pieters 1986). The Fe-oxidzed surface, which contains substantial hematite, shows absorptions at wavelengths shorter than 1.0 $\mu$m due to charge transfer absorptions for Fe$^{3+}$.

Signature narrow absorption features at NIR are characteristic of water ice as well as hydrated minerals. As shown in Figures \ref{rc} and \ref{FullCase}, water ice and hydrated minerals have high reflectivities at 1 $\mu$m and absorb strongly at longer wavelengths. The drop in reflectance at wavelengths longer than $\sim$2 $\mu$m is due to the presence of both the very strong OH stretch fundamental and overtone of the H$_2$O bend near 3 $\mu$m. We note that the amount of equivalent water in hydrated minerals is small ($<\sim$15\%), but even minor amounts of water (or OH) lead to prominent absorption features in the infrared. Sharp features due to bends and stretches of OH and H$_2$O occur at 1.4 and 1.9 $\mu$m in hydrated minerals and 1.5 and 2.0 $\mu$m in ice.

The non-linear effect of intimate mixing of different minerals can serve to subdue the absorptions of present phases. Darker phases are especially effective in hiding other constituents. Even though bright plagioclase is often the most abundant constituent in basalt, basalt is typically dark due to nonlinear mixing between plagioclase and dark constituents present at the few percent level. These nonlinear affects require careful analysis of detectability thresholds. Nevertheless, as discussed further in 4.2.1 and 4.2.2 below, the absorptions of ultramafic and hydrous surface types can be distinguished in broad-band telescopic data.

\begin{table}[htdp]
\caption{Average geometric albedo and color of an airless exoplanet fully covered by 8 types of crust listed in Table \ref{Assemblage} in NIR J, H and K bands. The two planet scenarios correspond to one case where reflection dominates the NIR planetary flux and the other case of close-in exoplanets where thermal emission extends to the NIR. Note that in the reflection-dominated case, the ultramafic surface is the only type of crust considered that has significantly higher albedo in the K band than in the J band, hydrous crusts are the only ones with lower albedos in the K band than J band, and ice-rich silicate surface is the only type of crust that has lower albedo in the H band than in the J band.}
\begin{center}
\begin{tabular}{l|lllll|lllll}
\hline
\hline
Planet Scenario & \multicolumn{5}{|l|}{G Star, 1 AU} & \multicolumn{5}{|l}{G Star, 0.15 AU} \\
\hline
Crust Type		&	J	&	H	&	K	& K-J & H-J & J & H & K & K-J & H-J \\
\hline
Metal-rich		&	0.12  &	0.18  &	0.19 & 0.07 & 0.06 & 0.17 & 0.56 & 2.28 & 2.12 &  0.40	\\
Ultramafic	&	0.32  &	0.55  &	0.61 & 0.29 & 0.23 & 0.33 & 0.59 &	0.78  & 0.45 &	0.25	\\
Feldspathic	&	0.56  &	0.61  &	0.63 & 0.07 & 0.05 & 0.56 & 0.62 &	0.68  & 0.12 &	0.06	\\
Basaltic		&	0.24  &	0.25  &	0.24 & 0.00 & 0.01 & 0.28 & 0.56 &	2.03  & 1.75 &	0.28	\\
Granitoid		&	0.60  &	0.65  &	0.71 & 0.11 & 0.05 & 0.60 & 0.65 &	0.72  & 0.13 &	0.06	\\
Clay			&	0.68  &	0.73  &	0.51 & -0.17 & 0.05 & 0.68 & 0.73 &	0.59  & -0.09 &	0.05	\\
Ice-rich silicate&	0.34  &	0.23  &	0.21 & -0.13 & -0.11 & 0.36 & 0.43 & 1.51  & 1.15 &	0.07		\\
Fe-oxidized	&	0.21  &	0.21  &	0.21	& 0.00 & 0.00 & 0.25 & 0.58 & 2.24  & 2.00 &	0.33	\\
\hline
\end{tabular}
\end{center}
\label{crust_bb}
\end{table}

\subsubsection{Broad-Band Spectral Feature of the Ultramafic Surface}

The NIR J band (1.1-1.4 $\mu$m) is sensitive to the detection of ultramafic crusts that contain ferrous igneous minerals on an exoplanet's surface. In the J band, surface absorbers include olivine and pyroxene (see Figure \ref{rc}). Made of these two types of minerals, the ultramafic crust leads to a distinctive J-band geometric albedo significantly lower than the H band and the K band (see Figure \ref{crust_scatter} and Table \ref{crust_bb}), unique in our set of representative surfaces. We define a key parameter for characterization of surface composition as the difference between the K-band geometric albedo ($A_{\rm g} ({\rm K})$) and the J-band geometric albedo ($A_{\rm g} ({\rm J})$). This parameter basically describes the ``color" of exoplanetary surfaces in the NIR wavelengths. Note that thermal emission may contribute to the apparent geometric albedo, as shown by Figure \ref{FullCase} and Table \ref{crust_bb}. By saying ``color" we refer to the true geometric albedo with component of thermal emission properly removed. For an exoplanet completely covered by ultramafic crust, $A_{\rm g} ({\rm K}) - A_{\rm g} ({\rm J}) = 0.29$, whereas all other crustal types give value less than 0.1 (see Table \ref{crust_bb}). It is therefore very likely that an ulframafic surface on exoplanets will stand out in the reflectance spectra.

Macroscopic mixtures between the ultramafic surface and other types of surfaces linearly lowers the spectral contrast of the J-band absorption feature. When mixed with other surfaces that have relatively flat spectra, such as basalt and granite, the J band feature of an ultramafic surface will appear to be shallower. As shown in Figure \ref{crustmix}, the key parameter $A_{\rm g} ({\rm K}) - A_{\rm g} ({\rm J})$ depends linearly on the percentage of ultramafic crust on the planet's surface. A detection of $A_{\rm g} ({\rm K}) - A_{\rm g} ({\rm J})>0.2$ indicates that more than a half of the planetary surface need to be covered by ultramatic materials.


\subsubsection{Broad-Band Spectral Feature of the Water and Hydrated Mineral Surfaces}

\begin{figure}[h]
\begin{center}
 \includegraphics[width=1\textwidth]{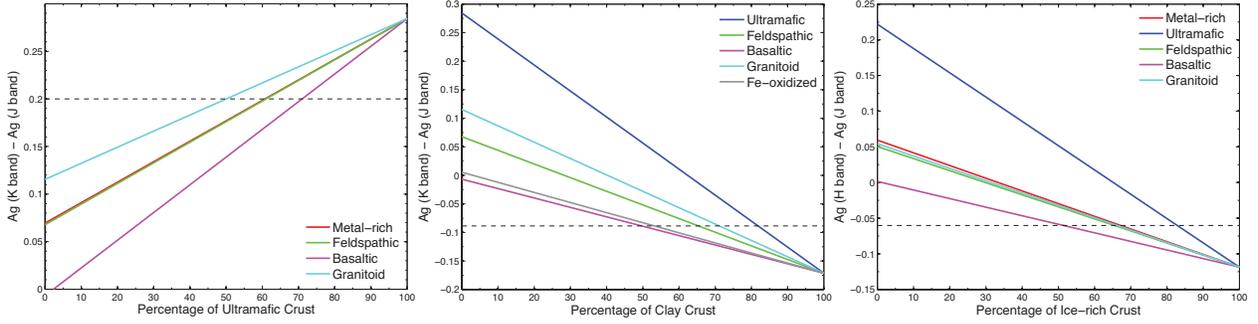}
 \caption{Colors of exoplanetary surfaces that contain macroscopic mixture of two types of crusts. The horizontal axis is the percentage of feature-creating constituents. The vertical axis the difference between the planet's geometric albedo in two NIR bands. The color of each line corresponds different companion constituents, tabulated as the legend in the plots.
 The horizontal lines are the limits of the color differences in order to infer that as least 50\% of the planetary surface is ultramafic, clay, and ice-rich, respectively. }
 \label{crustmix}
  \end{center}
\end{figure}

The NIR H band (1.5-1.8 $\mu$m) is suitable for the detection of water ice on an exoplanet's surface. At the H band, the most prominent absorber is water ice. In fact, water ice absorption in the H band is so strong that small amount (10\%) of water ice on a very reflective surface can significantly reduce the planetary geometric albedo. An ice-rich surface is the only type of surface in our sample that has lower albedo in H band than in J band (Table \ref{crust_bb}). As a result, if one observes a high J-band albedo and a low H-band albedo for an airless exoplanet, it is very likely that water ice exists on the planet's surface. If $A_{\rm g} ({\rm H}) - A_{\rm g} ({\rm J}) < -0.06$, more than half of the planetary surface is likely to be ice-rich (see Figure \ref{crustmix}).

The NIR K band (2-2.4 $\mu$m) is sensitive to water ice and hydrated minerals on an exoplanet's surface. We find that ice-rich surfaces and aqueously altered surfaces produce strong K-band absorption, which leads to a K-band albedo smaller than the J-band albedo (see Figure \ref{crust_scatter}). For example, for a generally reflective surface in K band, e.g., plagioclase- or olivine-rich, a small amount of water ice (10\%) can reduce the planetary geometric albedo significantly. As shown in Figure \ref{crustmix}, the key parameter $A_{\rm g} ({\rm K}) - A_{\rm g} ({\rm J})$ depends linearly on the percentage of clay crust on the planet's surface. A detection of $A_{\rm g} ({\rm K}) - A_{\rm g} ({\rm J})<-0.09$ indicates that more than a half of the planetary surface need to be covered by hydrated materials.

\section{Discussion}

\subsection{Effects of Atmospheres}

If the exoplanet has a thin atmosphere, molecular NIR absorptions will introduce additional features in its reflection and thermal emission spectra (e.g. Mars reflectance spectra in Figure \ref{SS}). Common molecules that actively absorb at NIR in an rocky exoplanet's atmosphere include H$_2$O, CO$_2$, CO, CH$_4$, NH$_3$, SO$_2$, etc (e.g., Seager \& Deming 2010). Based on the HITRAN molecular absorption line database, for an atmosphere of Earth-like temperature and pressure, 18 parts-per-million (ppm) of H$_2$O or 173 ppm of CH$_4$ can produce an integrated optical depth larger than unity in the J band (Hu \& Seager, in preparation). Similarly, 32 ppm of CH$_4$ can produce an integrated optical depth larger than unity in the H band; and 104 ppm of CO$_2$, or 4 ppm of CH$_4$, or 5 ppm of NH$_3$ can produce an integrated optical depth larger than unity in the K band (Hu \& Seager, in preparation).

For the broad-band infrared photometry, the effect of atmospheres on characterization of surface composition can be very serious. For example, we have demonstrated that water ice on the planet's surface produce H-band absorption, or negative $A_{\rm g} ({\rm H}) - A_{\rm g} ({\rm J})$. However, water vapor in the planet's atmosphere absorbs in J band, which will reduce the apparent contrast of the surface-related H-band feature. For another example, a J-band absorption feature in reflectance spectra can be interpreted as absorption of olivine and pyroxene on the surface, or as the absorption of water vapor in the atmosphere. Without any prior knowledge on the planet's atmosphere, it could be hard to draw conclusive surface compositions from broad-band spectrophotometry observations.

There are two possible ways to break the surface-atmosphere degeneracy in the exoplanet reflection spectra. First, observing the primary transit can determine the atmospheric composition of the exoplanet via transmission spectroscopy (e.g., Seager \& Deming 2010). The surface minerals do not induce any spectral features in the transmission of stellar radiation observable in primary transit. The difficulty of this approach is that the magnitude of atmospheric signal from the primary transit is proportional to the atmospheric scale height (e.g., Miller-Ricci et al. 2009; Seager 2010), therefore for a terrestrial-like atmosphere the signal may be too low to be detected. Second, high-resolution spectroscopy, rather than spectrophotometry, can distinguish a surface absorption from an atmospheric absorption. For reflection, the iron-related absorption in silicates are typically very wide and smooth in wavelength (see Figure \ref{FullCase}), but the atmospheric molecular features are typically narrow and only occur at specific wavelengths; for thermal emission, the silicate vibrational features have particular shapes which can hardly be produced by any plausible temperature structures in the planetary atmosphere (see Figure \ref{FullCase} and Figure \ref{Kepler20f} for an example of Kepler-20f). As a result, one can break the surface-atmosphere degeneracy in the exoplanet reflection spectra by achieving high spectral resolution.

In the extreme case, the refractory rocks may even be vaporized on the dayside and form a metal vapor atmosphere, as suggested for CoRot-7b by Schaefer \& Fegley (2010) and L\'eger et al. (2011). It has been shown that metal vapor atmospheres create narrow and strong metal transition lines in the visible wavelengths (e.g., Seager \& Sasselov 2000). It is therefore unlikely that metal vapor atmospheres impede spectral characterization of the surface beneath.

\subsection{Effects of Space Weathering}

\label{SpaceWeathering}

The surface of an airless planet is subject to space weathering. The processes of space weathering include collision of galactic cosmic rays, sputtering from solar wind particles, bombardment of micrometeorites, etc. Space weathering will produce a thin layer of nanophase iron on mineral grain surfaces and agglutinates in the regolith, altering surface spectral properties significantly. In general, as a surface matures due to space weathering, it becomes darker, redder, and the depth of its diagnostic absorption bands is reduced (Pieters et al. 1993; Hapke 2001).

\begin{figure}[h]
\begin{center}
 \includegraphics[width=0.8\textwidth]{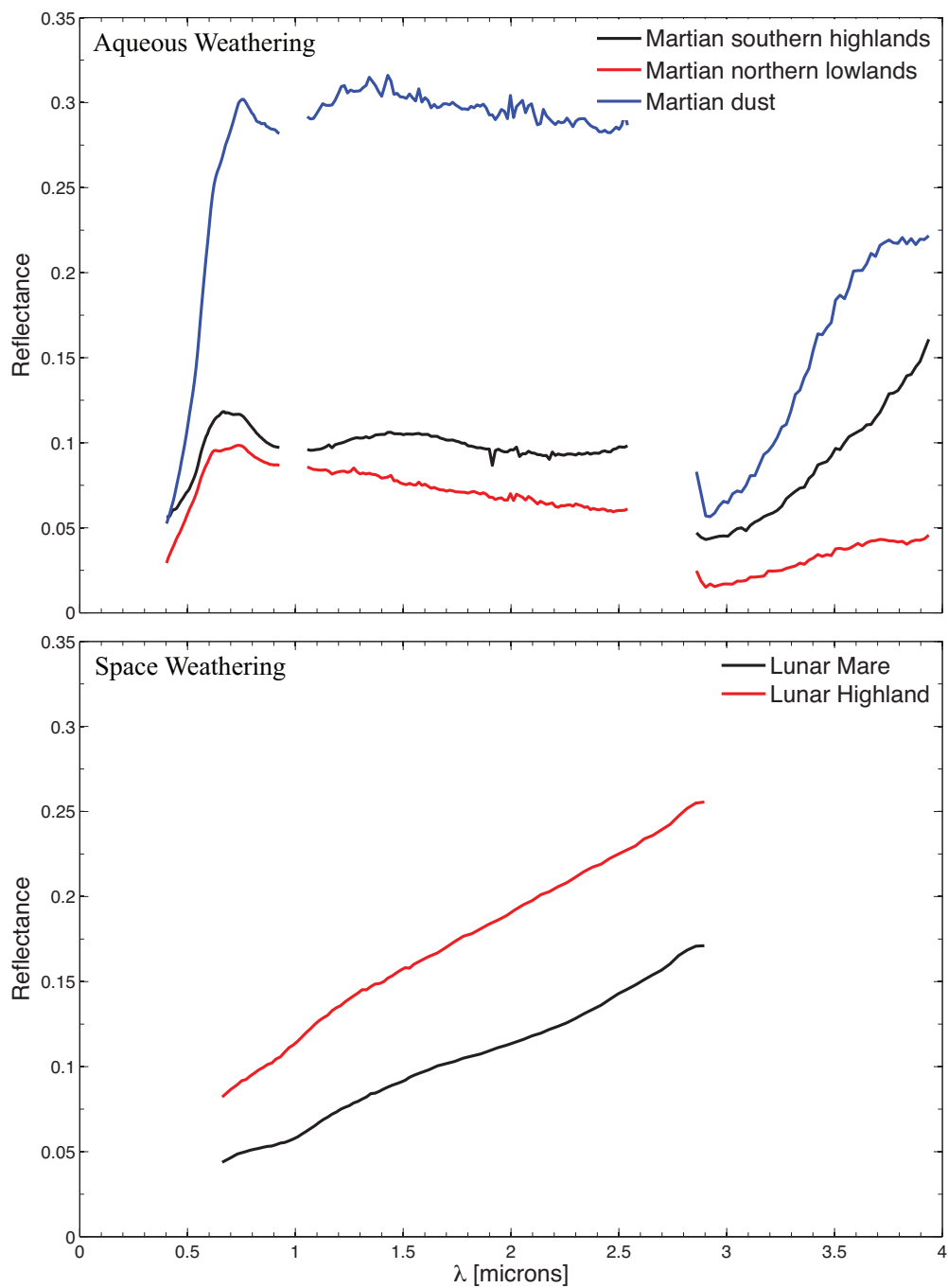}
 \caption{Reflectance spectra of weathered surfaces on the Moon and Mars.
 Mars spectra are measured by the OMEGA Visible and Infrared Mineralogical Mapping Spectrometer onboard the Mars Express orbiter (Skok et al. 2010 for martian lowlands; Mustard et al. 2005 for martian highlands and martian dust). Gaps in the spectra correspond to detector boundaries.
 The Moon's spectra are measured by the Moon Mineralogy Mapper (Isaacson, personal communication).
 The drop in reflectivity at $\sim2.8$ $\mu$m in all the martian spectra is due to strong OH-stretch fundamentals and the presence of water ice.
 }
 \label{MoonMars}
  \end{center}
\end{figure}

The broad-band diagnostic features considered in this paper will be subdue for an exoplanet covered by space-weathered (mature) surfaces. We present several reflectance spectra of selected mature lunar and martian surfaces in Figure \ref{MoonMars}. Fe$^{2+}$ absorption features at 1 $\mu$m and 2 $\mu$m become very shallow for the mature lunar mare (e.g., Figure \ref{MoonMars} vs. Figure \ref{rc}). We see that space weathering is very effective in reducing the diagnostic features. Nonetheless, various mechanisms can refresh a planet's surface in the course of evolution, which include impact cratering, volcanism, plate tectonics, etc. As a result, the fact that space weathering reduces the 1-$\mu$m and 2-$\mu$m spectral feature makes this detection indicative of active or recent re-surfacing on the planet.

Finally, for weathered surfaces the slope of reflectance spectra could be broadly diagnostic of the composition of weathering products and the nature of weathering. As shown in Figure \ref{MoonMars}, after space weathering, the reflectance spectrum from the basaltic lunar mare have a steep upward slope, as expected. In contrast, the basaltic martian lowlands become bluer after weathering. The downward slope in the spectra of mature martian lowlands likely indicates silica and nanophase iron oxides coatings caused by chemical reactions in thin films of water on basaltic rocks (Mustard et al., 2005). Although weathering reduces the contrast of spectral features, it causes slopes in reflectance spectra relevant to understanding planetary surface processes.

\subsection{Detection Potential}

\begin{figure}[h]
\begin{center}
 \includegraphics[width=0.8\textwidth]{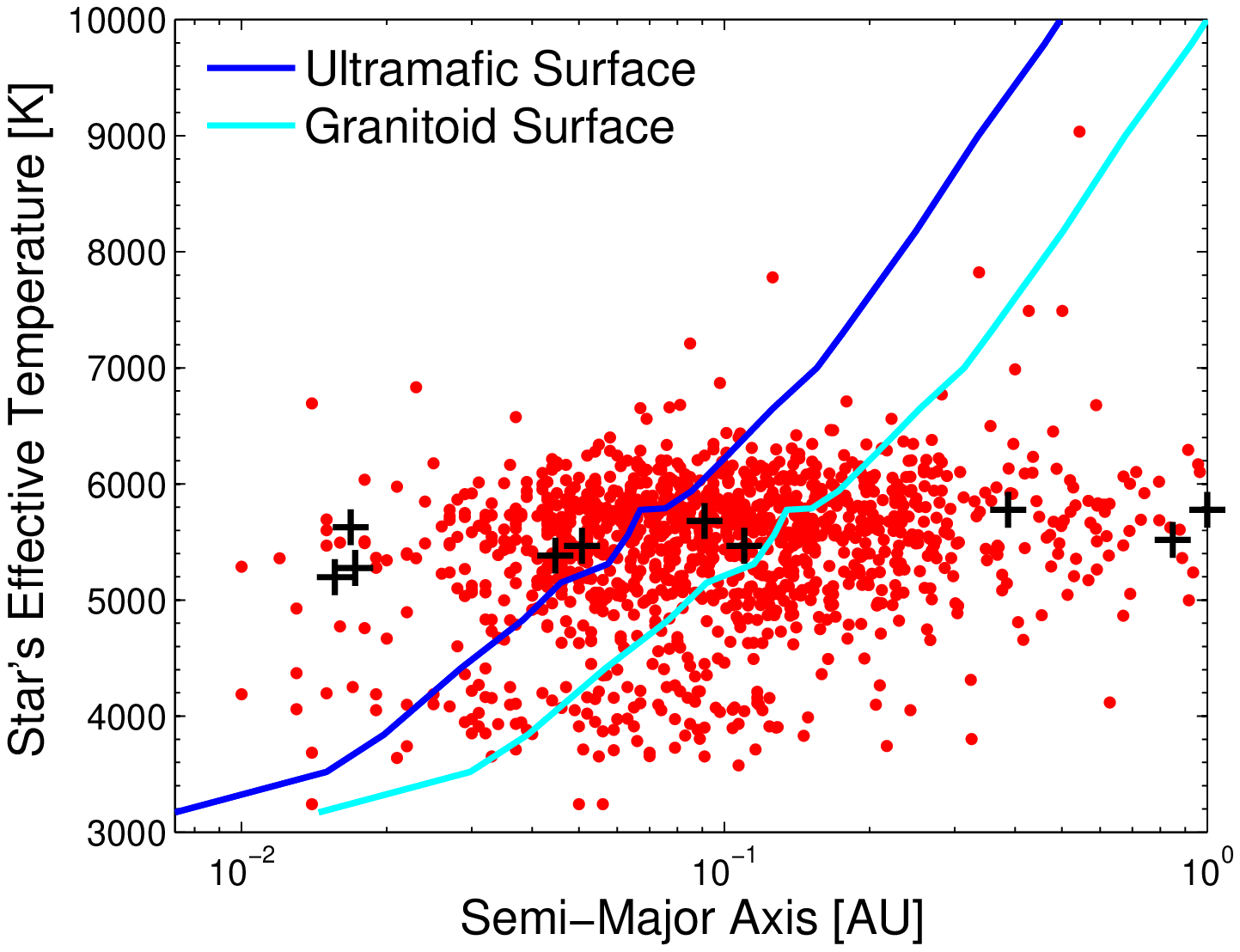}
 \caption{
 Relations between the spectral type of main-sequence stars (represented by effective temperatures) and the innermost orbital distance for a rocky planet to stay unmelted.
  Red dots are Kepler-released planet candidates as of April 2011 (Borucki et al. 2011). From left to right, the black markers correspond to planets that have or may have rocky surfaces: 55 Cnc e, Kepler-10 b, Corot-7 b, Kepler-18 b, Kepler-20 e, Kepler-11 b, Kepler-20 f, Mercury, Kepler-22 b and Earth.
 The lines are obtained by comparing sub-stellar (hottest) temperature and the zero-pressure melting temperature of silicates. Wavelength-dependent reflectivity and emissivity are considered in the computation of substellar temperature. The two lines correspond to limiting cases of silicates, for which ultramafic is the most refractive with a melting temperature of 1600 K and granitoid is the least refractive with a melting temperature of 1000 K. The melting temperatures assumed are typical, but exact numbers sensitively depend on content of volatiles and detailed mixtures of minerals (e.g., Best \& Christiansen 2001).
A rocky planet that falls on the left side of the blue line certainly has molten lava on its surface, and a rocky planet that falls on the right side of the cyan line has solid silicates surface, i.e., the focus of this paper.
 }
 \label{SubStellar}
  \end{center}
\end{figure}

The characterization of airless rocky exoplanets' solid surfaces is not inherently more difficult than the characterization of their atmospheres. Via the thermal emission, the silicate, iron-oxide and metal-rich surfaces manifest themselves differently in the planetary spectra, with a spectral contrast comparable to that of atmospheric CO$_2$ or O$_3$ features. The secondary transit signal of an exoplanet is larger if the planet is closer to its host star. As shown in Figure \ref{SubStellar}, however, an airless rocky planet cannot be too close to the star in order to keep its surface solid, which limits the potential occultation depth of the planet. For an Earth-sized planet around a G type star, the occultation depth can be larger than 10 ppm in MIR, and the surface characterization by transits requires a photometric precision of 2 ppm in MIR (see Figure \ref{Kepler20f} for an example of Kepler-20f). Such precision and corresponding signal-to-noise ratio may be attainable by the James Webb Space Telescope (JWST) if the telescope would observe all transits in its 5-year nominal mission time (Belu et al. 2011). Much higher secondary transit depths may be possible if the rocky exoplanets are discovered to orbit around M dwarfs, because M dwarfs have smaller radius and surface temperature compared to solar-type stars. M dwarfs are the most feasible targets for the characterization of rocky exoplanets' atmospheres (e.g., Seager \& Deming 2010), the same for the characterization of rocky exoplanet's surfaces. For instance, an Earth-sized planet at a 0.02-AU orbit of an M dwarf with effective temperature of 3000 K and size of 0.5 $R_{\sun}$ can have solid silicate surfaces (Figure \ref{SubStellar}), and the secondary transit depth ranges from 50 to 100 ppm in the mid-infrared. Such planets, if discovered, may be proven to have certain type of silicate surfaces (e.g., ultramafic versus granitoid) or iron-oxide surfaces by MIR spectroscopy.

The possibility of rocky super-Earths should not be neglected. A number of transiting exoplanets with size of about 2 $R_{\earth}$ have been detected, and the mass and radius constraints cannot exclude the possibility of a silicate-dominant composition (e.g., Lissauer et al. 2011; Gautier et al. 2012; Borucki et al. 2012). Notably, Kepler-11b and Kepler-22b may have solid silicate surfaces according to their orbital semi-major axis (Figure \ref{SubStellar}). The secondary transit depths are 20 - 40 ppm for Kepler-11b and 5 - 10 ppm for Kepler-22b in the mid-infrared. As an application of our theoretical model, the contrast of the silicate features is computed to be 10 ppm in the 7 - 13 $\mu$m band for Kepler-11b and 2 ppm in the 15 - 25 $\mu$m band for Kepler-22b. Furthermore, large planets around M dwarfs, although not yet discovered, have the best potential for surface characterization. For example, a 2-$R_{\earth}$ planet orbiting at 0.02 AU around a 0.5-$R_{\sun}$ M star having effective temperature of 3000 K, the secondary transit depth is up to $\sim$10 parts-per-million (ppm) at VNIR ($\lambda<3$ $\mu$m), and up to 300 ppm in MIR. The planet's thermal emission flux at $\sim10$ $\mu$m can vary between 200 and 300 ppm for different types of crust. As a comparison, the warm SPITZER has achieved the photometric precision as high as 65 ppm for bright stars such as 55 Cnc (e.g. Demory et al. 2011). Characterization of rocky exoplanets' surfaces via the thermal emission is therefore possible with current spacebased facilities if suitable targets are discovered.

Surface characterization by VNIR reflectance of rocky exoplanets is beyond the reach of current observation technology. The occultation depth of secondary eclipse at VNIR is fundamentally limited by the melting temperature of silicate rocks. To maintain a solid surface, the exoplanet's surface temperature should be lower than the melting temperature. Averaging the emissivity over the stellar-radiation wavelengths and the thermal-emission wavelengths, the local energy balance equation for surface temperature (Eq. \ref{LocalBalance}) can be solved to give the planet's sub-stellar temperature ($T_{\rm p}$) as
\begin{equation}
T_{\rm p} = T_{*} \bigg(\frac{\epsilon_{VNIR}}{\epsilon_{MIR}}\bigg)^{1/4} \bigg(\frac{R_*}{D_{\rm p}}\bigg)^{1/2} \ ,\label{Tsurf}
\end{equation}
where $\epsilon_{VNIR}$ and $\epsilon_{MIR}$ are the Planck mean emissivity in the wavelength ranges of stellar and planetary emission, respectively. Algebra from equation (\ref{Tsurf}) and equation(\ref{AG}) gives the following expression of the occultation depth:
\begin{equation}
\frac{F_{\rm p}}{F_*} = A_{\rm g} \frac{\epsilon_{VNIR}}{\epsilon_{MIR}} \bigg(\frac{R_{\rm p}}{R_{\rm *}}\bigg)^{2} \bigg(\frac{T_{\rm p}}{T_*}\bigg)^4\ . \label{occo}
\end{equation}
Without considering the thermal emission, $A_{\rm g}$ is the true geometric albedo of the planet and is in the order of unity. We see that the occultation depth increases rapidly with the surface temperature and the radius of the planet. For an Earth-like exoplanet around a Sun-like star, having surface temperature of 1000 K, the secondary occultation depth is about 0.1 ppm. Even for close-in rocky planets around M dwarfs (see Figure \ref{FullCase}), the VNIR reflection only leads to a transit depth in the order of 10 ppm. To characterize surface compositions by measuring secondary transits, the required photometric precision is in the order of 1 ppm. Photometric precision is fundamentally limited by the photon noise and the stellar variability. For Kepler observing a star of V mag of 13, the photometric precision per 6.5-hour transit is about 20 ppm (Koch et al. 2010). Therefore, the required photometric precision for surface characterization by VNIR reflectance is currently not possible. The photometric precision might be significantly improved by accumulating a large number of transits, with spacebased broad-band photometry instruments observing certain key objects for long periods, which requires a comprehensive understanding of noises, systematics, and stellar variability.

In the future, direct imaging can be suitable for characterizing solid surfaces of exoplanets. If direct imaging could spatially resolve the exoplanetary system, long-cadence observations can be carried out to obtain low-resolution spectra of the planet's reflection and thermal emission. Specific absorption features of mafic minerals, water ice and hydrated minerals in the reflection, as well as silicate and iron-oxide features in the thermal emission presented in this paper could stand out with broad-band photometry, and might be detectable by direct imaging.

\subsection{Connection of Surface Composition to the Planetary Interior and Evolution}

Planetary radius and mass inferred from primary transits and radial velocity measurements constrain the density of transiting exoplanets. The assemblage of minerals comprising the surface of a rocky planetary body provides additional valuable constraints on understanding exoplanet interior structure and geologic evolution. Planetary surface composition depends on the bulk composition of the exoplanet, the history and nature of magmatic and thermal processes, and the subsequent interaction of produced solids with atmospheric volatiles and/or the space environment.  We list as follows, in the order of detection likelihood, several type of planetary surfaces and their implications on the planet's interior structure and geological history.

A silicate surface, detectable via the prominent silicate features in the MIR thermal emission, will resolve the ambiguity of whether or not the planet has a significant envelope of volatiles. Due to the uncertainties in the measurements of mass and radius, the constraints of planetary interior are always ambiguous with various interpretation acceptable by data (e.g., Rogers \& Seager 2010). Theoretical studies of the volatile evolution may provide additional but indirect constraints (see Fressin et al. 2012 for an example of Kepler-20e and Kepler-20f). With the detection of planetary thermal emission in MIR and the identification of surface silicate features described in this paper, the possibility of significant volatile envelope can be readily excluded, and the planet can be confirmed to have silicate surface and mantle. The surface characterization enabled by MIR spectroscopy will therefore provide an essential dimension of constraints on the interpretation of mass-radius relationship of the planet.

A surface bright in VNIR, inferred from low equilibrium temperature, is likely to have felsic composition. Although the overall VNIR reflectivity is largely controlled by unknown factors such as surface roughness and weathering processes, no known surface process can increase the surface reflectivity. In other words, the detection of a bright surface indicates that the surface is intrinsically bright when fresh. As shown in Figure \ref{FullCase}, a bright surface can be feldspathic, granitoid, and clay; all are silica-rich (termed ``felsic"). A felsic surface on a planet of size larger than Mars is probably produced by slow intrusion of molten lava, which indicates plate-tectonics and potential geological setting for the origins of life (Best \& Christiansen 2001; Southam et al. 2007). A caveat here is that high VNIR reflectivity can also be attributed to atmospheric effects, such as bright clouds and hazes. A careful study, probably with transmission spectra, needs to be carried out to distinguish a bright solid surface against a bright cloud deck.

An ultramafic surface indicated by the 1-$\mu$m absorption feature in reflectance spectra implies either mantle overturn of the planet or very high temperature lavas. For rocky planets such as Mars and Earth, mantle pressures lead to the retention of Al in garnet, making it unavailable for feldspar formation. Consequently, the predicted primary crust composition, following mantle overturn to form a stable density stratrification, is dominated by Mg-rich olivines and pyroxenes (Elkins-Tanton et al., 2005). As the surface matures, the 1-$\mu$m absorption feature diminishes. As a result, if strong 1-$\mu$m feature is identified from the J-band absorption, one may infer that the lava eruption or mantle overturn was geologically recent. The key indicator mineral is olivine and its strong contrast in the J-band due to the 1-$\mu$m absorptions makes detection possible.

A surface with hydrous materials, indicated by signature absorption features of ice or OH, implies substantial volatile inventory and constrains the planetary temperature to less than $\sim$700 K over geologic time to retain these materials on the surface. Hydrous materials, such as clays, would indicate liquid water having interacted with the crust, a parameter relevant to the habitability of the planet.

\section{Summary}

We have developed a theoretical framework to investigate reflection and thermal emission spectra of airless rocky exoplanets. We have modeled representative planetary surface types as fully covered by particulate mineral assemblages, whose mineral compositions depend on formation and evolutionary history of the planet. The most prominent spectral features and their geological implications are listed in the order of detectability.
\begin{itemize}
\item The silicate surface leads to Si-O vibrational features in both the 7 - 13 $\mu$m band and the 15 - 25 $\mu$m band. The silicate features are universal for all rocky surfaces, the magnitude of which can be as large as 20 K in terms of brightness temperature for an airless Earth analog. The silicate features allow unambiguous detection of rocky exoplanets via mid-infrared spectroscopy.
\item Iron-oxidation leads to Fe-O vibrational features in the 15 - 25 $\mu$m band, which indicates an oxidized surface geochemical environment.
\item The location of emissivity maxima at the short-wavelength edge (7 - 9 $\mu$m) of the main silicate feature uniquely indicates the silica content, which can be used to determine whether the surface is ultramafic, mafic or felsic. The location of the emissivity maxima may be found via photometry using 3 or more narrow-bands in the 7 - 9 $\mu$m region.
\item  The iron crystal-field electron band at 1 $\mu$m is indicative of olivine and pyroxene minerals. In terms of NIR broad-band photometry, the difference between the J-band geometric albedo and the K-band geometric albedo ($A_{\rm g}({\rm K})-A_{\rm g}({\rm J})$) may distinguish ultramafic surfaces with these minerals, implying either mantle overturn of the planet or very high temperature lavas.
\item The OH vibrational bands beyond 2 $\mu$m are indicative of either surface water ice or hydrated minerals, which indicates extant or past water on the planet's surface. A broad vibrational absorption band at 1.5 $\mu$m is diagnostic of water ice on the surface.
\end{itemize}

We propose that observations of rocky exoplanet reflection and thermal emission will provide valuable information on the planet's surface mineral composition. Broad-band photometry of secondary eclipses in MIR may be able to identify a planetary surface of silicates or iron oxides (i.e., a rocky surface) via prominent spectral features. The required photometry precision is 2 ppm for planets around G stars and 20 ppm for planets around M stars. Next-generation space infrared facilities, such as the JWST, will likely be able to identify silicate surfaces around Sun-like stars. Reflected stellar light can also be used to uniquely determine the mineral composition of an exoplanet's surface, and is particularly useful for identifying ferrous or hydrated surface compositions. The occultation depth of the secondary eclipse in VNIR, however, is very small and beyond the reach of current technology. The required photometric precision for surface characterization via reflection is less than 0.1 part per million for an Earth-sized exoplanet around a Sun-like star, and 1 part per million for an Earth-sized exoplanet around an M dwarf. Although reflection of Earth-sized rocky planets are difficult to observe, planets as large as 55 Cnc e, if orbiting around M dwarfs, are best targets for long-period spacebased photometric monitoring. Eventually, the unique identification of minerals on exoplanets' surfaces may rely upon direct imaging. There are degeneracies among surface-related absorption features and atmospheric features, which may be broken by achieving high spectral resolution or by observing the primary transit and obtaining the transmission spectra. The surface mineral composition provides important constraints on the composition and the geological history of the exoplanet, which will constitute a new dimension of exoplanet characterization.

\acknowledgments

Thanks to P. Isaacson for providing M$^3$ lunar spectra and W. Calvin for providing modeled water ice spectra. RH is supported by NASA Earth and Space Science Fellowship (NESSF/NNX11AP47H).

\clearpage

\clearpage

\clearpage

\clearpage

\clearpage

\clearpage

\clearpage

\clearpage

\end{document}